\documentclass[lettersize,journal]{IEEEtran}
\usepackage{amsmath,amsfonts}
\usepackage{algorithmic}
\usepackage{array}
\usepackage[caption=false,font=normalsize,labelfont=sf,textfont=sf]{subfig}
\usepackage{textcomp}
\usepackage{stfloats}
\usepackage{url}
\usepackage{verbatim}
\usepackage{graphicx}

\usepackage{multirow}
\usepackage{multicol}
\usepackage{tcolorbox}
\usepackage{lipsum} 
\usepackage{booktabs}
\usepackage{adjustbox}

\usepackage{multirow}
\usepackage{makecell}
\usepackage{amsmath,amssymb,amsfonts}
\usepackage{array}   
\usepackage{multicol}
\usepackage{hyperref}
\usepackage{enumitem}
\usepackage{caption}
\usepackage{tabularx}
\usepackage{dsfont}

\def\BibTeX{{\rm B\kern-.05em{\sc i\kern-.025em b}\kern-.08em
    T\kern-.1667em\lower.7ex\hbox{E}\kern-.125emX}}
\usepackage{balance}
\begin{document}
\title{Tackling the Modality Inhibitance Effect with Wavelet-based Modality Integration in Multi-modal Recommendation}

\author{Rongqing Kenneth Ong and Andy W.~H. Khong, \IEEEmembership{Member, IEEE}%
\thanks{Rongqing Kenneth Ong and Andy W.~H. Khong are with the School of Electrical and Electronic Engineering at Nanyang Technological University. Andy W.~H. Khong also holds a joint appointment with the Lee Kong Chian School of Medicine in the same university.}
}

\markboth{IEEE Transactions on Knowledge and Data Engineering}%
{How to Use the IEEEtran \LaTeX \ Templates}

\maketitle

\begin{abstract}
Harnessing rich modality features has recently gained prominence in multi-modal recommendation systems (MRSs). To delineate user modality preferences, existing works focus on modality fusion, which can be achieved via attentive summation or concatenation. 
In this work, we introduce the mobility inhibitance effect\textemdash a phenomenon which limits existing MRSs when these models transit from dual-modal to tri-modal configurations resulting in significant performance degradation.
In particular, apart from the modal-specific contamination encapsulated in the newly introduced modality, the inclusion of new features results in excessive informational redundancy, thereby severely impeding effective integration within current approaches. To address the mobility inhibitance effect, we propose a new multi-scale modality wavelet (MMWave) multi-modal recommender, which aims to fully capitalize on essential modality features and suppress modal-specific noise. This is acheived by employing wavelet decomposition from a multi-scale (time \& frequency) perspective, ensuring that only essential (non-redundant) features of each modality are extracted. We then account for the latent associations between item modality features by designing a new structure-aware multi-modal graph learning module that captures high-order latent structural information pertaining to items. 
To modulate the relative contributions of modality features and achieve effective integration, we then propose a new adaptive modality regulator component that employs a new cross-weight integration strategy. Experiments conducted on four real-world datasets validate the efficacy of our proposed model. The source code for this work has been made publicly available at~\url{https://github.com/kennethorq/MMWave}.
\end{abstract}


\section{Introduction}
In the dynamic realm of e-commerce platforms, recommendation systems play an indispensable role in assisting users discover items of interest. To meet such demands, many systems rely on collaborative filtering techniques, which leverage user-item historical interactions to capture users’ patterns of interest. However, due to the rapid increase in multimedia content~\cite{ma2024leveraging,li2024multimodal,zhang2025survey, zhang2023beyond} comprising images, textual data, and videos, many traditional collaborative filtering frameworks face the new challenge of incorporating diverse modalities effectively. This has become a core focus within the research domain of multi-modal recommender systems (MRSs).

In the existing literature, conventional MRSs focus on modality fusion as the principal means of modeling user preferences across modalities. Pioneering works, including VBPR~\cite{he2016vbpr} and Deepstyle~\cite{liu2017deepstyle}, explore visual modality fusion by projecting visual features into a lower-dimensional space and subsequently integrating them with item-identity (ID) embeddings via concatenation and element-wise summation. Since user-item interactions can be structured intrinsically as a bipartite graph, researchers have increasingly adopted graph neural networks~\cite{he2020lightgcn, dai2022towards, zhao2022joint, caisimple} to encode the high-order behavioral and multi-modal semantic information into both the user and item encoding process. For instance, MMGCN~\cite{wei2019mmgcn} projects the heterogeneous modality inputs into multiple user-item views and characterises user interests via the concatenation of the learned modality embeddings. As an extension, GRCN~\cite{wei2020graph} uses multi-modal signals to refine the user–item view by suppressing erroneous links arising from behavioral interactions. To validate the benefits of latent structures between items for item representation learning, LATTICE~\cite{zhang2021mining} first constructs a modality-specific graph that reflects different semantic viewpoints of the item. An attention mechanism is then used to determine the relative importance of each modality within the fusion process. More recently, FREEDOM~\cite{zhou2023tale} demonstrated that trainable latent graphs are not critical and, therefore, are frozen prior to training, improving the overall computational and memory efficiency of the model. The LGMRec~\cite{guo2024lgmrec} was subsequently proposed to explore the use of hypergraphs for discerning users' global interests while complementing local graph-based embeddings, thereby generating a more robust representation of user preferences.

Although existing studies have demonstrated some success in exploiting multiple modalities, these modality signals are integrated indiscriminately, allowing 
redundant cross-modal signals to permeate the item representation learning process. This coarse-grained integration causes excessive informational redundancy, resulting in inconsistent performance across modalities\textemdash a limitation that becomes even more pronounced when additional channels are incorporated. To illustrate the above, we adopt the real-world Tiktok dataset~\cite{wei2023multi}, consisting of three distinct modalities (e.g., visual, text, audio) for our experiments. We performed experiments under varying input conditions and evaluated three recent state-of-the-art fusion based models: MMGCN~\cite{wei2019mmgcn}, FREEDOM~\cite{zhou2023tale}, and LGMRec~\cite{guo2024lgmrec} for comparison. As shown in Table~\ref{tab: case study}, it is evident that these fusion-based models suffer from poor performance when additional modalities are incorporated. While modality features can improve the performance of recommender systems, these models exhibit suboptimal performance in the presence of visual, text, and audio modalities. More importantly, these results indicate that different models excel with different modality combinations\textemdash FREEDOM performs best with text and audio, whereas MMGCN and LGMRec achieve ideal performance with visual and text. This observation underscores the ineffectiveness of existing MRSs in fully harnessing the potential of each modality's unique features, signaling the persistence of redundant cross-modal information across multiple modalities.

Table~\ref{tab: case study} also shows that the non-fusion BM3 model achieves optimal results when all modalities are utilized. Since modality imbalance occurs in fusion- \emph{and} non-fusion-based models~\cite{peng2022balanced, liang2022mind}, the above results imply that even with a negligible degree of modality imbalance within a dataset, fusion-based models still suffer from poor performance. We attribute this poor performance to a different and largely unresolved challenge: the modality inhibitance effect~(MIE) caused by \emph{redundant cross-modal} signals arising from multiple modality integration. 
Addressing the MIE by fully utilizing the various modalities is not effective since each modality may be contaminated with its own distinct type of noise. To harness the potential of each modality effectively, a well-designed MRS must be meticulously formulated to extract essential features from within the contaminated modality data and to strategically address the MIE arising from the integration of multiple modalities.

\begin{table}[t]
\caption{Performance of existing fusion and non-fusion based Multi-modal Recommenders with varying modalities on Tiktok Dataset.}
    \centering
    \begin{tabularx}{\columnwidth}{llXXXX}
        \toprule
        Non-fusion based & Modality & R@5 & R@10 & N@5 & N@10 \\
        \midrule
        \multirow{2}{*}{BM3~\cite{zhou2023bootstrap}} 
        & Visual \& Text & 0.0397 & 0.0693 & 0.0259 & 0.0354 \\
        & Visual \& Audio & 0.0420 & 0.0673 & 0.0269 & 0.0349 \\
        \multirow{2}{*}{(WWW'23)} 
        & Text \& Audio & 0.0417 & 0.0709 & 0.0262 & 0.0355 \\
        & Full & \textbf{0.0437} & \textbf{0.0722}  & \textbf{0.0287} & \textbf{0.0379} \\
        \midrule
        Fusion based & Modality & R@5 & R@10 & N@5 & N@10 \\
        \midrule
        \multirow{2}{*}{MMGCN~\cite{wei2019mmgcn}} 
        & Visual \& Text & 0.0299 & \textbf{0.0578} & 0.0199 & \textbf{0.0288} \\
        & Visual \& Audio & \textbf{0.0364} & 0.0427 & \textbf{0.0219} & 0.0241 \\
        \multirow{2}{*}{(MM'19)} 
        & Text \& Audio & 0.0154 & 0.0486 & 0.0094 & 0.0202 \\
        & Full & 0.0309 & 0.0542 & 0.0152 & 0.0223 \\
        \midrule
        \multirow{2}{*}{FREEDOM~\cite{zhou2023tale}} 
        & Visual \& Text & 0.0394 & \textbf{0.0614} & 0.0252 & 0.0321 \\
        & Visual \& Audio & 0.0397 & 0.0607 & 0.0256 & 0.0322 \\
        \multirow{2}{*}{(MM'23)} 
        & Text \& Audio & \textbf{0.0424} & 0.0581 & \textbf{0.0294} & \textbf{0.0344} \\
        & Full & 0.0394 & 0.0571 & 0.0262 & 0.0319 \\
        \midrule
        \multirow{2}{*}{LGMRec~\cite{guo2024lgmrec}} 
        & Visual \& Text & \textbf{0.0322} & \textbf{0.0588} & 0.0198 & 0.0282 \\
        & Visual \& Audio & 0.0295 & 0.0512 & 0.0193 & 0.0262 \\
        \multirow{2}{*}{(AAAI'24)} 
        & Text \& Audio & 0.0312 & 0.0568 & \textbf{0.0208} & \textbf{0.0290} \\
        & Full & 0.0305 & 0.0502 & 0.0204 & 0.0267 \\
        \bottomrule
    \end{tabularx}
    \label{tab: case study}
\end{table}
In this work, we focus on the distillation of essential modality features while jointly suppressing modal-specific contamination. Leveraging well-established principles in signal processing~\cite{mallat1989theory}, we draw inspiration from the concept of wavelet transformation, which has been shown to be effective in modality denoising~\cite{li2020wavelet, salama2024semantic, oka2025wavelet}. Through wavelet transformation, modality signals are decomposed into multiple levels of granularities, with each set of wavelet coefficients capturing progressively finer or coarser details of the modality features. This decomposition enables high-level multi-resolution analysis~\cite{zhang2019wavelet}, thereby facilitating both temporal and spatial localization that are critical for effective processing of sequential (e.g, text, audio) and spatial (e.g, image) modalities. Moreover, the frequency-time localization property of wavelet transformation facilitates the suppression of dominant high-frequency components typically associated with noise~\cite{zhang2024multimodal}, thus allowing the extraction of most essential features through strategically guided attenuation. Motivated by the ability of multi-resolution analysis to facilitate robust feature extraction, we propose a new multi-scale modality wavelet (MMWave) recommender system designed to extract essential features and effectively integrate diverse modalities of varying types. In essence, MMWave comprises three key components: 1) wavelet modality mixer, 2) structure-aware multi-modal graph learning, and 3) adaptive modality regulator.

To ensure that essential modality features are extracted, MMWave performs wavelet-based modality decomposition by breaking down each modality into sub-bands of features comprising both temporal and frequency characteristics. Since noise contamination often occurs in the form of high-frequency components~\cite{al2021towards, polspoel2022importance}, the proposed model capitalizes on the single low-frequency component within the localized time-frequency representation for modality signal attenuation. In this regard, we introduce an efficient yet effective parametric filter to modulate the essential signals that approximate the original modality features. The uni-modal features are then reconstructed only with the modulated modality signals, ensuring that the most essential and representative features are captured. 

To account for both collaborative and semantically affiliated information, we design a new structure-aware multi-modal graph learning module to encode high-order behavioral and structural characteristics pertaining to both users and items. Finally, to mitigate MIE and achieve adaptive integration, we formulate a new adaptive modality regulator module that derives fused representations from the modulated uni-modal signals before using them to assign reciprocal weights across modalities. This process allows the model to capture complementary information with greater precision. Experiments on four real-world datasets show the efficacy of the proposed model. The contributions of this work are threefold:
\begin{itemize}
    \item We introduce the modality inhibitance effect as a key limitation in current multi-modal recommenders and propose a new wavelet-based model to mitigate its impact;
    \item We design a structure-aware multi-modal graph learning module to encode high-order behavioral and semantically correlated structural signals;
    \item We formulate a new adaptive modality regulator to enhance modality integration via cross-weighting between uni and fused representations, guided by behavioral injection.
\end{itemize}

\begin{figure*}[t]
    \centering
    \includegraphics[width=1\textwidth, height=0.5\textheight, keepaspectratio]{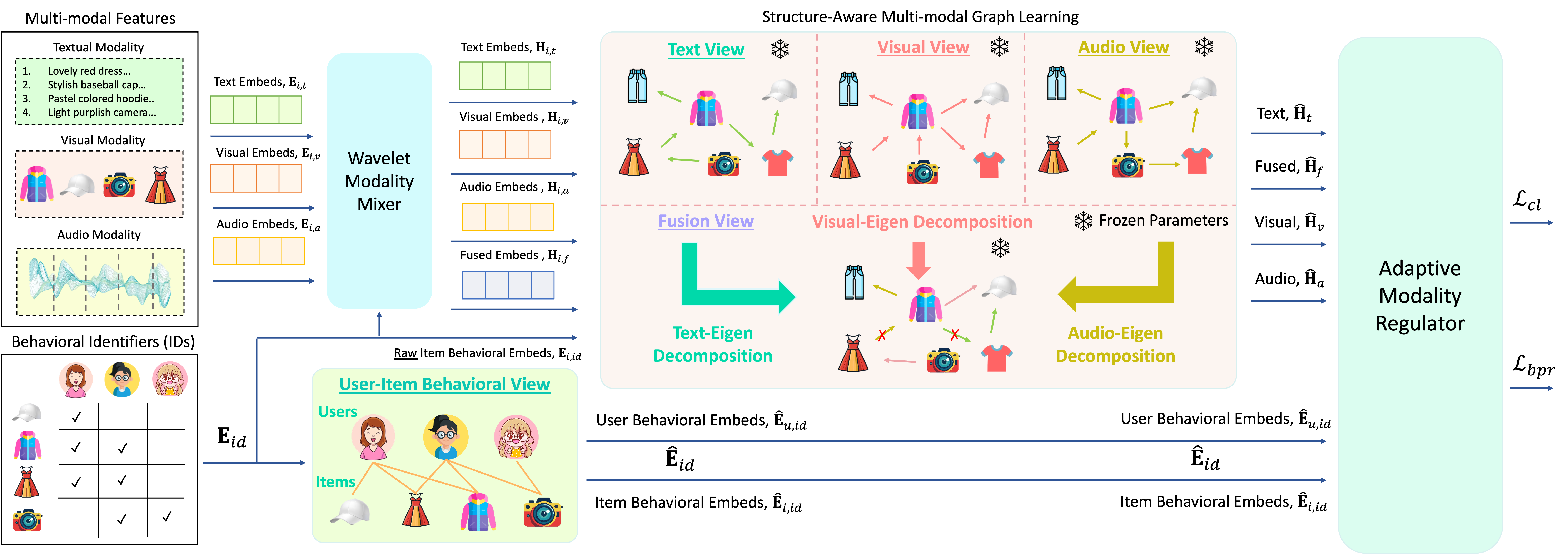}
    \caption{An overview of the proposed architecture highlighting its three core components: (i) wavelet modality filter, (ii) structure-aware multi-modal graph learning and (iii) adaptive modality regulator module.}
    \label{MMWave Architecture}
\end{figure*}

\section{Task Formulation and Review of Existing MRS}
\subsection{Task Formulation}
\label{Task Formulation}
Users' implicit interactions are first modeled by defining~$\mathcal{U} = \{ u_j \mid 1 \leq j \leq P \}$ and $\mathcal{I} = \{ i_k \mid 1 \leq k \leq Q \}$ as the user and item sets, where $P$ and $Q$ represent the respective cardinalities. Thereafter, an interaction matrix is denoted as $\mathbf{Y}\in\mathbb{R}^{P\times Q}$, where $y \in \{0,1\}$, with 1 signifying an observed interaction, and 0 indicating otherwise. To embed users and items, we initialise an ID embedding matrix $\mathbf{E}_{id}\in\mathbb{R}^{d\times\left( |\mathcal{U}| + |\mathcal{I}|\right)}$, with $d$ denoting the dimensionality of the embedding space. For each item $i$, we further define $\mathbf{e}_i^m \in \mathbb{R}^{d_m}$ as its corresponding modality feature vector. Here, $d_m$ indicates the feature dimension with respect to modality $m$, where $m$ belongs to the modality set $\mathcal{M}$. It is important to note that in this work, we focus on three modalities, represented as $\mathcal{M} = \{v,t,a\}$, with $v$, $t$ and $a$ signifying visual, text, and audio modalities, respectively. Given the interaction data and the multi-modal features of items, the goal of MRSs is to accurately model user preferences by estimating the interaction probability $\widehat{y}_{ui}$ between user $u$ and item $i$.

\subsection{Multi-modal Recommendation System}
The exploitation of multiple modalities~\cite{pomo2025recommender, zhou2025does} has fueled the advancement of MRSs, which enables a deeper understanding of users' fine-grained modality preferences. Conventionally, pre-trained neural networks such as VGG-16~\cite{simonyan2014very} for visual input and the Sentence Transformer~\cite{reimers2019sentence} for textual input have been utilized for the extraction of multi-modal features. These features serve as auxiliary information to MRSs. For instance, the classical VBPR~\cite{he2016vbpr} unimodal approach makes use of visual modality features by first concatenating with its item ID embeddings before applying a multi-layer perceptron to project the embeddings into the latent space for representation learning. To model users' varying degrees of preferences, MAML~\cite{liu2019user} adopts a two-layer, attentional feedforward network on the visual and textual representations of the items. Following a similar line of work, VECF~\cite{chen2019personalized} models intricate user preferences by leveraging pre-segmented image patches and an attention mechanism to identify important regions of the image.

\subsection{Graph-based Multi-modal Recommenders}
Since user-item interactions can be inherently represented as a bipartite graph~\cite{he2020lightgcn, caisimple, mao2021ultragcn, wang2019neural, zhou2023bootstrap, xu2025mentor, tang2025r2mr, ong2023quad}, recent works began leveraging graph neural networks to capture high-order complex relationships between users and items. For instance, LATTICE~\cite{zhang2021mining} accounts for essential item-item latent connections by first constructing similarity graphs that are distinct to each modality. The fusion of these modal-specific graphs is then guided by importance scores\textemdash these scores are used to combine the graphs prior to applying graph convolution for generating the final fused representation. Similar to redundancies in adaptive filtering~\cite{khong2005, khong2006stereophonic, khong2008algorithms}, redundancies that exist in graph learning have also been reported in~\cite{zhou2023tale}, which highlights the inefficiency of learning graphs during training. In this aspect, FREEDOM~\cite{zhou2023tale} freezes the graph structure while effectively capturing important relationships between the modalities of item-item. More recently, LGMRec~\cite{guo2024lgmrec} adopts hypergraph learning by treating attributes as hyperedges to capture global structural dependencies between items and attributes, thereby modeling users’ global interests. CMDL~\cite{lin2025contrastive} captures modality-specific and shared information through modality disentangled learning, using contrastive regularization to separate modality-specific and invariant representations. While majority of the works have shown promising results with dual modalities (e.g., image, text), their performance deteriorates significantly as more modalities are incorporated, exposing the modality inhibitance effect and underscoring the complexity of effective modality integration. In this work, we focus on mitigating the modality inhibitance effect through effective feature extraction and robust modality integration. Unlike prior approaches, the proposed wavelet-based design, MMWave, decomposes modality features into multiple levels of granularity, enabling finer-grained feature extraction and more precise cross-modal integration.

\section{The Proposed MMWave Multi-Modal Recommender}
Figure~\ref{MMWave Architecture} illustrates the proposed MMWave model comprising three core components: (i) wavelet modality mixer that effectively extracts essential modality features, (ii) structure-aware multi-modal graph learning that encodes semantically associated behavioral and structural cues from user-item and item-item viewpoints, and (iii) adaptive modality regulator module that captures intricate cross-modal relationships and performs dynamic integrations of diverse modalities.
\begin{figure*}[t]
    \centering
    \includegraphics[width=1\textwidth, height=0.5\textheight, keepaspectratio]{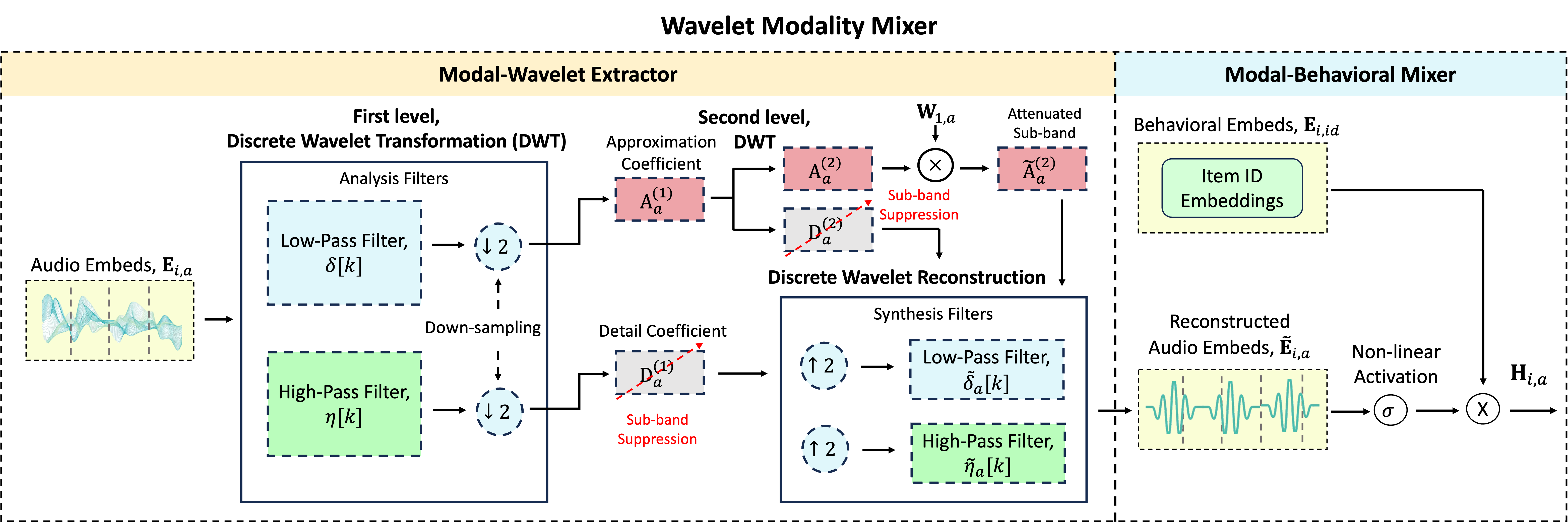}
    \caption{Proposed wavelet modality mixer module, comprising two key stages: (i) modal-wavelet extractor and (ii) modal-behavioral mixer. For illustration purposes, only the audio modality is presented in this example.}
    \label{WMM Module}
\end{figure*}

\subsection{Wavelet Modality Mixer}
Modality features convey rich, expressive information that is essential for capturing the nuances of user preferences. However, in real-world applications such as e-commerce~\cite{zhou2023comprehensive}, redundant information is often embedded within modalities, 
thereby hindering effective modality integration, resulting in MIE occurrence. Resolving this phenomenon necessitates a two-stage process: (i) the extraction of core informative features and (ii) the formulation of a coherent modality integration framework. As an initial step, a common practice in existing works~\cite{zhou2023bootstrap,zhang2022latent,zhou2023tale,ong2025spectrum} is to project the modality features into a shared latent feature space prior to performing modality integration~\cite{zhou2023tale} or denoising~\cite{yu2023multi}. However, such approaches inevitably introduce noise by entangling informative modality signals with noise in a compressed representation space. 

In contrast with the above approaches and one that performs spectral shaping~\cite{lin2012}, we directly extract essential modality features from the raw input modalities, as illustrated in Fig.~\ref{WMM Module}. As opposed to MLP-based denoising approaches, which rely on learning a parametric mapping from noisy (modality) inputs to clean outputs, the wavelet transform in MMWave offers an explicit multi-resolution decomposition, thereby enabling more precise extraction of distinct feature information. Applying $L$ levels of discrete wavelet transformation, we first obtain a hierarchical set of subband wavelet coefficients 
\begin{equation}
\mathcal{W}^{(L)}_{m}(\mathbf{E}_{i,m}) = \left\{ \mathbf{A}_m^{(L)},\, \mathbf{D}_m^{(l)} \,\middle|\, l = 1, \dots, L \right\}
\label{eq:1}
\end{equation}
by recursively decomposing the modality features such that
\begin{align}
a^{(l)}_{m}[n] &= \sum_k \delta[k] \cdot \mathbf{e}^{(l-1)}_{m}[2n - k], \qquad \delta[k] = \left[ \frac{1}{\sqrt{2}}, \frac{1}{\sqrt{2}} \right] \\
d^{(l)}_{m}[n] &= \sum_k \eta[k] \cdot \mathbf{e}^{(l-1)}_{m}[2n - k], \qquad \eta[k] = \left[ -\frac{1}{\sqrt{2}}, \frac{1}{\sqrt{2}} \right]
\end{align}
correspond, respectively, to the approximation and detail coefficients of the modality features. Here, $\mathbf{A}_m^{(L)} \in\mathbb{R}^{(d_m/2^{L}) \times |\mathcal{I}|}$ and $\mathbf{D}_m^{(l)} \in\mathbb{R}^{(d_m/2^{l}) \times |\mathcal{I}|}$ are the approximate and detailed coefficient matrices, respectively. The variables $\delta[k]$ and $\eta[k]$ are the low- and high-pass analysis filters of the Haar wavelets~\cite{mallat1989theory}, $\mathbf{e}^{(l-1)}_{m}$ denotes the modality inputs at the $L-1$-th decomposition level, with the indexing term $2n-k$ reflecting a stride-2 downsampling convolution. Although many alternative wavelet bases~\cite{daubechies1988orthonormal, cohen1992biorthogonal,meyer1992wavelets} are available, we adopt the Haar wavelets as the basis function due to its computational efficiency and strong localization properties~\cite{mallat1989theory}.

With the low-frequency (approximation) and high-frequency (detail) subbands, we propose an efficient filtering mechanism that modulates the most informative approximation signals of each modality such that the filtered modality signal is given by
\begin{equation}
\mathbf{\widetilde{A}}_{m}^{(L)} = f_{m} \left( \mathbf{A}_{m}^{(L)} \right) = \mathbf{W}_{1,m} \odot \mathbf{A}_{m}^{(L)},
\end{equation}
where $f_{m}$ denotes the modal-specific filter function, the operator $\odot$ signifies the point-wise (Hadamard) product,  and the $(d_m/2^{L}) \times |\mathcal{I}|$ matrix $\mathbf{W}_{1,m}$ is the modal-specific trainable attenuation weights. Given that the approximation coefficient preserves the most informative features, this efficient yet effective strategy enhances the informative signals that define each modality's unique characteristics.

In domains such as image and natural language processing~\cite{chen2024adaptive,chen2025ambiguity, chaovalit2011discrete}, a straightforward approach to signal restoration~\cite{banham1996spatially, chang2000spatially} through feature denoising is to reconstruct the modality features using the entire set of acquired wavelet coefficients. However, since our objective is to extract the essential modality signals rather than to fully restore the original modality features, and given that modality noise is often associated with high-frequency components~\cite{al2021towards, polspoel2022importance, li2020wavelet}, we propose to reconstruct the modality features directly from the attenuated signals that preserve only the most essential features. To this end, we introduce a modified form of the inverse discrete wavelet transform given by
\begin{equation}
\mathcal{W}^{~-1}_{m} : \widetilde{\mathbf{a}}^{(l-1)}_{m}[k] = \sum_n \widetilde{\delta}[n] \cdot \widetilde{\mathbf{a}}^{(l)}_{m}\left[ \left\lfloor \frac{k - n}{2} \right\rfloor \right],
\end{equation}
\begin{equation}
\mathbf{\widetilde{E}}_{i,m} = \mathbf{\widetilde{A}}_{m}^{(0)} = \mathcal{W}^{-1} \left( \mathbf{\widetilde{A}}_{m}^{(L)}\right), \quad \mathbf{\widetilde{E}}_{i,m} \in \mathbb{R}^{n \times d},
\end{equation}
where $\widetilde{\delta}[n]$ is defined as the low-frequency synthesis filter of the Haar wavelet used for reconstruction, and the indexing term $(k-n) / 2$ arises from the inverse operation of stride-2 downsampling during the analysis phase. On this basis, redundant information is effectively filtered out, while only the most essential modality-specific signals are amplified and preserved, thereby ensuring discriminative and well-separated modality-specific representations. These essential modality features $\mathbf{\widehat{E}}_{i,m}$ are then projected onto a shared latent space using the multilayer perception (MLP) such that
\begin{equation}
\mathbf{\widehat{E}}_{i,m} = \mathbf{W}_{2,m} \mathbf{\widetilde{E}}_{i,m} + \mathbf{b}_{1,m},
\end{equation}
where $\mathbf{W}_{(.)}\in\mathbb{R}^{d\times d}$ and $\mathbf{b}_{(.)}\in\mathbb{R}^{d}$ are defined as the trainable modality weights and biases, respectively. Given that early-stage fusion produces joint representations that capture rich cross-modal dependencies~\cite{xu2025survey}, we construct an initial first-order fused representation by combining the attenuated modality features giving 
\begin{equation}
\widehat{\mathbf{E}}_{i,f} = \sum_{m \in \mathcal{M}} \alpha_{m} \, \widehat{\mathbf{E}}_{i,m}
\end{equation}
where the subscript $f$ denotes the fused representation and $\alpha_{m}$ denotes the attention scores with respect to the relative importance of each modality. 

Similar to~\cite{yu2023multi, ong2025spectrum} and to ensure behavioral information is well preserved within each modality, we extract both modality-specific $\mathbf{H}_{i,m}$ and fused preference $\mathbf{H}_{i,f}$ signals under the guidance of behavioral signals such that
\small
\begin{align}
\mathbf{H}_{i,m} = &\theta_{gate}^{m}\left( \mathbf{E}_{i,id}, \widehat{\mathbf{E}}_{i,m} \right) = \mathbf{E}_{i,id} \odot \sigma \left( \mathbf{W}_{3,m} \widehat{\mathbf{E}}_{i,m}  \right), \label{eq:uni-modal features}\\
\mathbf{H}_{i,f} = &\theta_{gate}^{f}\left( \mathbf{E}_{i,id}, \widehat{\mathbf{E}}_{i,f} \right) = \mathbf{E}_{i,id} \odot \sigma \left( \mathbf{W}_{4,f} \widehat{\mathbf{E}}_{i,f} \right). \label{eq:fused-modal features}
\end{align}
\normalsize
Here, $\theta_{gate}^{(\cdot)}$ signifies the modality gating function used for injection of behavioral features, while $\mathbf{E}_{i,id}$ represents the item ID embeddings. $\mathbf{W}_{(.)}\in\mathbb{R}^{d\times d}$ denotes the trainable weights, and $\sigma$ being the non-linear sigmoid function. We shall illustrate, in Subsection.~\ref{final_subsection}, how the wavelet transformation effectively extracts essential signals, laying the foundation for the subsequent stages of our proposed model.

\subsection{Structure-aware Multi-modal Graph Learning}
\subsubsection{Item-item multi-modal views}
With the extraction of the uni-modal $\mathbf{H}_{i,m}$ and fused features $\mathbf{H}_{i,f}$, this module involves encoding the high-order behavioral and structural multi-modal features. As shown in~\cite{zhou2023tale, yu2023multi, zhou2023enhancing}, propagating both collaborative and semantically associated signals through graph convolutional network can improve the performance of a MRS. We therefore construct a set of modality-specific similarity matrices $\mathbf{S}_m$, one for each modality $m$. Each matrix encodes pairwise item affinities, with element $s_{a,b}^m$ in $\mathbf{S}_m$ denoting the degree of similarity between items row $a$ and column $b$ under modality $m$ such that
\begin{equation}
\mathbf{s}_{a,b}^m = \frac{\left( \mathbf{e}_a^m \right)^{\text{T}} \mathbf{e}_b^m}{\| \mathbf{e}_a^m \| \| \mathbf{e}_b^m \|}.
\end{equation}
Here, $\mathbf{e}^m_{(\cdot)} \in\mathbb{R}^{d_m}$ denotes the raw modality embedding vector. Similar to~\cite{zhang2021mining} and for each item, we retain the most informative signals from neighboring items by performing graph sparsification~\cite{chen2009fast}\textemdash preserving only the top $K$ edges with the highest similarity scores to its neighbors, i.e.,
\begin{equation}
\label{knn_sparsification}
\dot{s}_{a,b}^m =
\begin{cases} 
s_{a,b}^m, & s_{a,b}^m \in \text{top-}K_m(\{\mathbf{s}_{a,c}^m, \, c \in \mathcal{I}\}); \\
0, & \text{otherwise},
\end{cases}
\end{equation}
where $\dot{s}_{a,b}^m$ quantifies the top $K_m$ edge weights between items $a$ and $b$. To alleviate gradient instability issues arising from vanishing and exploding gradients~\cite{kipf2016semi}, the similarity matrix is subsequently normalized as
\begin{equation}
\ddot{\mathbf{\mathcal{S}}}_m = \mathbf{D}_m^{-1/2} \dot{\mathbf{\mathcal{S}}}_m \mathbf{D}_m^{-1/2},
\end{equation}
where $\mathbf{D}_m^{-1/2}$ corresponds to the degree matrix of $\dot{\mathbf{\mathcal{S}}}_m$.

While constructing the fusion graph using simple heuristic rules is possible (e.g., pre-defined attention weights~\cite{zhang2021mining,zhou2023bootstrap}), such an approach overlooks the structural information embedded in the modality-specific graph, including both global structure and local neighbourhood patterns. Beyond neglecting structural information, these approaches also suffer from poor scalability in higher-order multi-modal settings, as the number of fusion hyperparameters increases linearly with the number of modalities and consequently imposes substantial tuning overhead. To bridge this gap, we propose a new paradigm for learning fusion graph structures by first performing spectral decomposition on the acquired uni-modal $m$ graphs $\ddot{\mathbf{\mathcal{S}}}_m$ such that
\begin{equation}
\mathbf{V}_m \, \boldsymbol{\Lambda}_m \, \mathbf{V}_m^\top = \ddot{\mathbf{\mathcal{S}}}_m.
\end{equation}
Here, $\mathbf{V}_m \in \mathbb{R}^{N \times N}$ contains the orthonormal eigenvectors (i.e., $\mathbf{V}_m^\top \mathbf{V}_m = \mathbf{I}$), and $\boldsymbol{\Lambda}_m = \operatorname{diag}(\lambda_{1,m}, \dots, \lambda_{N,m})$ with eigenvalues being sorted in descending order. To capture complementary modality structural cues, we refine and reconstruct the uni-modal graphs using eigenvalues derived from other modalities as
\begin{equation}
\widehat{\mathbf{\mathcal{S}}}_m  = \frac{1}{|\mathcal{M}| - 1} \sum_{\substack{m' \in \mathcal{M} \\ m' \ne m}} \ddot{\mathbf{\mathcal{S}}}_m  \odot \widehat{\boldsymbol{\lambda}}^{m'}.
\end{equation}
With the uni-modal graphs enriched by structural information from its corresponding modality, we generate a structure-aware modality fusion graph using average pooling
\begin{equation}
\widehat{\mathbf{\mathcal{S}}}_f  = \frac{1}{|\mathcal{M}|} \sum_{m\in\mathcal{M}} \widehat{\mathbf{\mathcal{S}}}_m, \quad \widehat{\mathbf{\mathcal{S}}}_f \in \mathbb{R}^{NxN}.
\end{equation}
Here, the fusion graph serves as a unified representation that encapsulates the diverse structural characteristics of each modality, while ensuring equitable structural contribution through eigenvalue-based weighting. It is important to note that the uni-modal and fusion graphs are pre-constructed prior to training, which incurs no additional computational or memory overhead during training. 

Having established both the uni-modal $\ddot{\mathbf{\mathcal{S}}}_m$ and structural infused $\widehat{\mathbf{\mathcal{S}}}_f$ graphs, the first-order uni-modal and fusion features derived from (\ref{eq:uni-modal features}) and~(\ref{eq:fused-modal features}) can be effectively distilled into high-order representations through an efficient light graph convolutional layer~\cite{he2020lightgcn}, formulated as
\begin{equation}
\widehat{\mathbf{H}}_{i,m}^{(q)} = \widehat{\mathbf{\mathcal{S}}}_m \mathbf{H}_{i,m}^{(q-1)}, \qquad \widehat{\mathbf{H}}_{i,f}^{(q)} = \ddot{\mathbf{\mathcal{S}}}_f \mathbf{H}_{i,f}^{(q-1)}.
\label{n_item_mod}
\end{equation}
where $\mathbf{H}_{i,m}^{(q-1)}$ and $\mathbf{H}_{i,f}^{(q-1)}$ denote, respectively, the uni-modal and fused item representations at the $(q-1)$th layer. Following the same principle, we can propagate high-order users' uni-modal and fused features recursively through a weighted sum aggregation layer such that
\begin{eqnarray}
\widehat{\mathbf{h}}_{u,m}^{(q)} &= \sum_{i \in \mathcal{N}_u} \frac{1}{\sqrt{|\mathcal{N}_u| |\mathcal{N}_i|}} \widehat{\mathbf{h}}_{i,m}^{(q-1)}, 
\\
\qquad \widehat{\mathbf{h}}_{u,f}^{(q)} &= \sum_{i \in \mathcal{N}_u} \frac{1}{\sqrt{|\mathcal{N}_u| |\mathcal{N}_i|}} \widehat{\mathbf{h}}_{i,f}^{(q-1)},
\label{n_user_mod}
\end{eqnarray}
where $\mathcal{N}_u$ and $\mathcal{N}_i$ denote the sets of neighbors for user $u$ and item $i$, respectively, and the term $1 / \sqrt{\scalebox{0.9}{$|\mathcal{N}_u| |\mathcal{N}_i|$}}$ is the symmetric normalization term introduced to mitigate the risk of overscaling.
Unlike existing approaches that employ shallow propagation (e.g., $q$=1) to avoid the introduction of latent noise in deeper layers, our proposed model achieves optimal performance at a slightly higher propagation depth by encoding key modality signals from semantically correlated neighbors, resulting in an effective transmission of only essential signals across multi-modal graph neighbourhoods.

\subsubsection{User-item behavior views}
This view focuses on capturing high-order collaborative signals derived from user interaction data. Since collaborative signals play a critical role in characterizing user behavioral patterns~\cite{he2020lightgcn, mao2021ultragcn,zhou2023tale, caisimple}, we perform recursive message propagation over the user-item interaction graph
\begin{equation}
\mathbf{E}_{id}^{(p)} = (\mathbf{D}^{-1/2} \mathbf{X} \mathbf{D}^{-1/2}) \: \mathbf{E}_{id}^{(p-1)}, \quad \mathbf{X} = \begin{bmatrix} 0 & \mathbf{Y} \\ \mathbf{Y}^\top & 0 \end{bmatrix},
\end{equation}
where $\mathbf{E}_{id}^{(p-1)}$ denotes both the user and item ID embeddings of the previous layer, and $\mathbf{D}^{-1/2}$ denotes the inverse square root of the diagonal degree matrix associated with the adjacency matrix $\mathbf{X}$. Using mean pooling as the aggregation function, we obtain the overall user and item ID embeddings via aggregation as follows
\begin{equation}
\widehat{\mathbf{E}}_{id} = \frac{1}{P+1} \sum_{i=0}^{P} \mathbf{E}_{id}^{(i)}, \qquad \widehat{\mathbf{E}}_{id} \in \mathbb{R}^{d \times (|\mathcal{U}|+|\mathcal{I}|)}.
\label{n_behav_propagation}
\end{equation}

\subsection{Adaptive Modality Regulator}
Central to addressing the modality inhibitance effect lies in the effective integration of high-order signals extracted from both uni-modal and fused features. Since the fused features encapsulate abundant essential information, integrating it with uni-modal representations requires a carefully designed strategy to better capture intricate cross-modal relationships\textemdash a limitation that is inadequately addressed by widely adopted methods such as attentive summation~\cite{guo2024lgmrec}. 
As such, we propose an efficient integration strategy that cross-weighs the uni-modal and fused features, ensuring substantial and adaptive contributions from each modality. We first define uni-modal and fused attention weights as
\begin{eqnarray}
\begin{aligned}
        &\gamma_{m} \;=\; \operatorname{norm}\!\left( \mathbf{z}_{m}^{\top}\, \tanh\!\bigl( \mathbf{W}_{5,m}\,\widehat{\mathbf{H}}_{i,m} + \mathbf{b}_{2,m} \bigr) \right),\\
        &\gamma_{f} \;=\; \operatorname{norm}\!\left( \mathbf{z}_{f}^{\top}\, \tanh\!\bigl( \mathbf{W}_{6,f}\,\widehat{\mathbf{H}}_{i,f} + \mathbf{b}_{3,f} \bigr) \right),
\end{aligned}
\end{eqnarray}
where $\mathbf{W}_{(.)}\in\mathbb{R}^{d\times d}$ and $\mathbf{b}_{(.)}\in\mathbb{R}^{d}$ denotes, respectively, the trainable modality weights and biases, and $\mathbf{z}_{(.)} \in \mathbb{R}^d$ are the uni-modal and fused attention vectors. To facilitate cross-guided modality contribution, we utilize the trainable fused attention weights extracted from fused embeddings to weigh all uni-modal features, i.e., 
\begin{equation}
\ddot{\mathbf{H}}_{i,m} = \gamma_f \widehat{\mathbf{H}}_{i,m},
\end{equation}
where $\ddot{\mathbf{H}}_{i,m}$ is the refined uni-modal features, adaptively re-weighted by the fused embeddings to suppress dominant biases and promote complementary integration. Conversely, we leverage the trainable uni-modal attention weights to re-weigh the fused embeddings such that 
\begin{equation}
\ddot{\mathbf{H}}_{i,f} = \frac{1}{|M|} \sum_{m \in M} \gamma_m \widehat{\mathbf{H}}_{i,f}
\end{equation}
is the refined fused features guided by the respective uni-modal signals. This approach facilitates balanced modality integration and is inherently scalable, ensuring effectiveness as more modalities are incorporated.

Since behavioral signals are essential for understanding user preferences across uni-modal and fused representations, we leverage high-order collaborative signals to model modality preferences using a parametric gate function such that uni-modal and fused collaborative attention weights are, respectively, given by
\begin{eqnarray}
\begin{aligned}
        &\mathbf{R}_m = \chi_{pref}^{m} \left( \widehat{\mathbf{E}}_{id} \right) = \sigma(\mathbf{W}_{7,m} \widehat{\mathbf{E}}_{id} ),\\
        &\mathbf{R}_f = \chi_{pref}^{f} \left( \widehat{\mathbf{E}}_{id} \right) = \sigma(\mathbf{W}_{8,f} \widehat{\mathbf{E}}_{id} ),
\end{aligned}
\end{eqnarray}
where $\chi_{pref}^{(\cdot)}$ represents the uni-modal and fused gate functions, and $\sigma$ is defined as the non-linear sigmoid function. Finally, 
we apply mean aggregation over the adaptively refined uni-modal and fused features to obtain the overall multi-modal side features
\begin{equation}
\mathbf{H}_{s} = \frac{1}{|M|} \left( \sum_{m \in M} \ddot{\mathbf{H}}_{i,m} \odot \mathbf{R}_m \right) +  \left( \ddot{\mathbf{H}}_f \odot \mathbf{R}_{f} \right).
\end{equation}
In terms of optimization and to enhance representational discrimination between high-order behavioral signals and modality-side features, we utilize the InfoNCE~\cite{oord2018representation} contrastive loss to jointly supervise user and item representations such that loss associated with the users and items are defined, respectively, as
\begin{eqnarray}
\mathcal{L}_{cl}^{u} &= \sum_{u \in \mathcal{U}} -\log \frac{\exp(\widehat{e}_{u,id} \cdot h_{u,s} / \tau)}{\sum_{v \in \mathcal{U}} \exp(\widehat{e}_{v,id} \cdot h_{v,s} / \tau)},
\label{cl_eq}
\\
\mathcal{L}_{cl}^{i} &= \sum_{i \in \mathcal{I}} -\log \frac{\exp(\widehat{e}_{i,id} \cdot h_{i,s} / \tau)}{\sum_{v \in \mathcal{I}} \exp(\widehat{e}_{v,id} \cdot h_{v,s} / \tau)}.
\label{cl_eq}
\end{eqnarray}
where $\tau$ is a hyperparameter that governs the sharpness of the similarity distribution. Incorporating the individual loss components, the final contrastive loss is obtained as $\mathcal{L}_{cl}=\mathcal{L}_{cl}^{u} + \mathcal{L}_{cl}^{i}$.

\subsection{Model Prediction and Optimization}
With the high-order behavioral $\widehat{\mathbf{E}}_{id}$ and modality side features $\mathbf{H}_{s}$, the resultant user and item representations are computed, respectively, as
\begin{equation}
\mathbf{e}_u^{*} = \widehat{\mathbf{e}}_{u,id} + \mathbf{h}_{u,s}, \quad \mathbf{e}_i^{*} = \widehat{\mathbf{e}}_{i,id} + \mathbf{h}_{i,s}.
\end{equation}
Under the full-sort prediction protocol, user preferences are estimated via the interaction probability $\widehat{y}(u,i) = \mathbf{e}_u^{*\top} \mathbf{e}_i$. For optimizing the proposed model, we leverage the BPR~\cite{rendle2012bpr} loss to reconstruct historical interactions by assigning higher scores to items with observed user feedback given as
\begin{equation}
\mathcal{L}_{bpr}=\sum_{(u,i,j)\in \mathcal{O}} -\mathrm{ln} \space \sigma\Big(\widehat{y}_{u,i}-\widehat{y}_{u,j}\Big).
\label{bpr_loss}
\end{equation}
Here, the interaction set $\mathcal{O}$ comprises triplets $(u, i, j)$ such that $(u, i)$ is the observed interaction from $\mathcal{O}^+$, whereas $(u, j)$ is the unobserved one from $\mathcal{O}^-$. The sigmoid function is denoted by $\sigma$. Thereafter, we jointly optimized the proposed model in parallel with the contrastive loss, yielding the overall loss function
\begin{equation}
\mathcal{L} = \mathcal{L}_{bpr} + \psi_{1} \mathcal{L}_{cl} + \psi_{2} ||\Theta||_{2}^{2},
\end{equation}
where $\psi_{(.)}$ denotes the hyperparameter governing the contribution of the contrastive task and L2 regularization loss.

\subsection{Distinguishing Modality Inhibitance Effect from Modality Imbalance}
In this section, we distinguish the modality inhibitance effect from modality imbalance~\cite{zhang2024multimodal, zhang2024modality}, a widely recognized issue in multi-modal learning, by first presenting their conceptual definitions, followed by their corresponding mathematical formulations. Modality imbalance occurs when a dominant and faster-converging modality overshadows its weaker counterparts. From the gradient optimization perspective, during multi-modal joint training, a dominant modality with faster convergence and higher performance can reduce the shared update step, thereby suppressing the optimization of other modalities. To formalize the preceding definitions, and based on the notations introduced in the task formulation in Section~\ref{Task Formulation}, the generalized form of the interaction probability of MRS can be defined as
\begin{equation}
\hat{y}_{ui} = f_{\theta}\left(\mathbf{e}_{u,id}, \mathbf{e}_{i,id}, \mathbf{p}_{u,m}, \mathbf{e}_{i,m}\right), 
\quad m \in \mathcal{M}
\end{equation}
where $f_{\theta}$ denotes the different approaches to model users' historical interactions and the multi-modal content pertaining to the items. The variable $\theta$ denotes all the trainable parameters of a given MRS model, while $\mathbf{p}_{u,m}$ and $\mathbf{e}_{i,m}$ are the user preference modality features and item modality features in general. For simplicity, we use the representative fusion-based model VBPR~\cite{he2016vbpr} as an illustrative example. VBPR performs feature fusion by concatenating ID embeddings (i.e., $\mathbf{E}_{u,id}$ and $\mathbf{E}_{i,id}$) with modality features such that 
\begin{align}
\bar{\mathbf{E}}_u &= [\mathbf{E}_{u,id} \,\Vert\, \mathbf{P}_{u,m}], \\
\bar{\mathbf{E}}_i &= [\mathbf{E}_{i,id} \,\Vert\, \hat{\mathbf{E}}_{i,m}].
\end{align}
Here, $\hat{\mathbf{E}}_{i,m} = \mathbf{W}_m \mathbf{E}_{i,m}$ denotes the item projected modality representation while $\mathbf{P}_{u,m}$ represents the user preference matrix associated with modality $m$. The predicted likelihood between user $u$ and item $i$ can, therefore, be re-written as
\begin{equation}
\hat{y}_{ui} = \mathbf{e}_{u,id}^{\top}\mathbf{e}_{i,id} 
+ \sum_{m \in \mathcal{M}} \mathbf{p}_{u,m}^{\top}\hat{\mathbf{e}}_{i,m},
\end{equation}
where $\mathbf{e}_{u,id}$ and $\mathbf{e}_{i,id}$ denote the ID-based user and item embeddings, respectively. The term $\mathbf{e}_{u,id}^{\top}\mathbf{e}_{i,id}$ captures the collaborative preference scores from user-item interactions. The vector $\mathbf{p}_{u,m}^{\top}$ denotes the preference representation of user $u$ towards modality $m$, and $\mathbf{p}_{u,m}^{\top}\hat{\mathbf{e}}_{i,m}$ determines the modality-specific preference score between the user and item under modality $m$. With reference to the BPR loss defined in Eq.~(\ref{bpr_loss}), for a training triplet $(u,i,j)$, where $i$ is a positive item and $j$ is a negative item for user $u$, we substitute the multi-modal score into the loss function and obtain
\begin{equation}
\mathcal{L}_{uij} = -\ln \sigma \left( \mathbf{e}_{u,id}^{\top}\mathbf{e}_{i,id} - \mathbf{e}_{u,id}^{\top}\mathbf{e}_{j,id} + \sum_{m \in \mathcal{M}} s_{uij}^{m} \right),
\end{equation}
where
\begin{equation}
s_{uij}^{m} = \mathbf{p}_{u,m}^{\top}\hat{\mathbf{e}}_{i,m}
-
\mathbf{p}_{u,m}^{\top}\hat{\mathbf{e}}_{j,m}
\end{equation}
denotes the ranking contribution of modality $m$. Similarly, since $\hat{\mathbf{e}}_{i,m} = \mathbf{W}_m \mathbf{e}_{i,m}$, we can re-write the equation in the form of
\begin{equation}
s_{uij}^{m} = \mathbf{p}_{u,m}^{\top}\mathbf{W}_{m}
\left(\mathbf{e}_{i,m} - \mathbf{e}_{j,m}\right).
\label{ranking contribution}
\end{equation}
Here, $s_{uij}^m$ quantifies the extent to which modality $m$ helps to distinguish the positive item $i$ from the negative item $j$ for user $u$. The total ranking margin can then be defined as
\begin{equation}
z_{uij} = \mathbf{e}_{u,id}^{\top} \mathbf{e}_{i,id} - \mathbf{e}_{u,id}^{\top} \mathbf{e}_{j,id} + \sum_{m \in \mathcal{M}} s_{uij}^{m}.
\end{equation}
The gradient magnitude, with respect to the modality-specific ranking contribution $s_{uij}^m$, is given by
\begin{equation}
\left| \frac{\partial \mathcal{L}_{ui}}{\partial s_{uij}^{m}} \right|
= \frac{1}{1 + \exp\left(z_{uij}\right)}.
\label{Partial Derivative of Loss}
\end{equation}
Since all modality-specific contributions $\{ s_{uij}^{m} \}_{m \in \mathcal{M}}$ are accumulated within the same ranking margin, this gradient term is shared across modalities. Therefore, for any given two modalities $m_1$ and $m_2 \in \mathcal{M}$, we have
\begin{equation}
\left| \frac{\partial \mathcal{L}_{uij}}{\partial s_{uij}^{m_1}} \right| = \left| \frac{\partial \mathcal{L}_{uij}}{\partial s_{uij}^{m_2}} \right|.
\end{equation}
Using gradient descent, the modality-specific trainable parameters are updated as 
\begin{equation}
\left( \mathbf{p}_{u,m}^{\top} \mathbf{W}_m \right)^{(t+1)} = \left( \mathbf{p}_{u,m}^{\top} \mathbf{W}_m \right)^{(t)} - \eta \frac{\partial \mathcal{L}_{uij}}{\partial s_{uij}^{m}} \left( \mathbf{e}_{i,m} - \mathbf{e}_{j,m} \right),
\end{equation}
where $t$ denotes the current optimization step, $(t+1)$ denotes the next optimization step, and $\eta$ is the learning rate of the optimizer. This formulation shows that, for the particular term $(\partial\mathcal{L}_{uij}/\partial s_{uij}^{m_1})$, when one modality $m_1$ dominates, it generates a high ranking contribution $\partial s_{uij}^{m_1}$. Specifically, the total ranking margin $z_{uij}$ becomes excessive, resulting in a large denominator in Eq.~(\ref{Partial Derivative of Loss}). The shared gradient term $1/(1+\text{exp}(z_{uij}))$ becomes smaller, thus reducing the update magnitude received by other modalities. Weaker modalities will therefore remain under-optimized. This phenomenon is known as the modality imbalance issue.

Unlike the modality imbalance issue, which focuses on unequal modality optimization, we define the modality inhibitance effect (MIE) as the inhibitory effect caused by \emph{redundant} modality signals arising from multiple modality integration. More specifically, MIE is not primarily concerned with whether individual modalities are unequally optimized. Instead, it captures whether integrating multiple modalities introduces redundant cross-modal signals that inhibit recommendation performance. Based on the above definition and to formalize MIE, we re-use the modality-specific ranking contribution $s_{uij}^m$ in Eq.~(\ref{ranking contribution}), since it directly reflects the recommendation signal provided by modality $m$. For any two modalities $m_1$ and $m_2 \in \mathcal{M}$, the cross-modal redundancy over a batch of training triplets $\mathcal{B}$ can be defined as
\begin{equation}
\mathcal{R}_{m_1,m_2} = \left| \operatorname{Corr}_{(u,i,j)\in\mathcal{B}} \left(s_{uij}^{m_1},s_{uij}^{m_2} \right) \right|.
\end{equation}
where $\operatorname{Corr}(\cdot)$ denotes the correlation function computed over the batch $\mathcal{B}$. A higher value of $\mathcal{R}_{m_1,m_2}$ implies that modalities $m_1$ and $m_2$ provide more similar ranking signals across the same training triples, suggesting stronger cross-modal redundancy. An intuitive analogy is that when two modalities, such as text and image, produce highly similar ranking signals, their integration may not introduce substantial new or complementary information. Instead, it primarily contributes to redundant cross-modal signals. For a modality subset $\mathcal{S} \subseteq \mathcal{M}$, the average pairwise cross-modal redundancy is defined as
\begin{equation}
\mathcal{R}(\mathcal{S}) = \frac{2}{|\mathcal{S}|(|\mathcal{S}|-1)} \sum_{\substack{m_1,m_2 \in \mathcal{S} \\ m_1 < m_2}} \mathcal{R}_{m_1,m_2}.
\end{equation}
where $|\mathcal{S}|$ denotes the number of modalities in $\mathcal{S}$, and the normalization term averages the redundancy over all modality pairs in $\mathcal{S}$. For brevity, we denote the recommender performance of a fusion-based multi-modal recommender using modality subset $\mathcal{S}$ as $\mathcal{P}_{\mathrm{MRS}}(\mathcal{S})$. Given the existing modality subset $\mathcal{S} \subseteq \mathcal{M}$, and an additional modality $m' \in \mathcal{M} \setminus \mathcal{S}$, the change in cross-modal redundancy after incorporating modality $m'$ is defined as
\begin{equation}
\Delta \mathcal{R}(m' \mid \mathcal{S}) = \mathcal{R}\left(\mathcal{S} \cup \{m'\}\right) - \mathcal{R}\left(\mathcal{S}\right).
\end{equation}
Similarly, the change in performance of MRS after incorporating modality $m'$ is defined as
\begin{equation}
\Delta \mathcal{P}_{\mathrm{MRS}}(m' \mid \mathcal{S}) = \mathcal{P}_{\mathrm{MRS}}(\mathcal{S} \cup \{m'\}) - \mathcal{P}_{\mathrm{MRS}}(\mathcal{S}).
\end{equation}
The MIE phenomenon occurs when the incorporation of a new modality increases the redundant cross-modal signals while degrading recommendation performance, specifically under the following conditions
\begin{equation}
\Delta \mathcal{R}(m' \mid \mathcal{S}) > 0
\quad \text{and} \quad
\Delta \mathcal{P}_{\mathrm{MRS}}(m' \mid \mathcal{S}) < 0.
\end{equation}
The above formulations delineate the conditions under which MI and MIE occur, thereby highlighting their fundamental differences.
\section{Experiments and Results}
We conducted a comprehensive set of experiments to investigate the following research questions:
\begin{itemize}
    \item \textbf{RQ1:} How effective is the proposed MMWave model when evaluated against state-of-the-art general and multi-modal recommenders?
    \item \textbf{RQ2:} How do the various components in MMWave contribute to its overall performance?
    \item \textbf{RQ3:} How does the robustness of the proposed model vary with hyperparameter changes?
    \item \textbf{RQ4:} How practical is the proposed MMWave in terms of computational cost and scalability?
    \item \textbf{RQ5:} To what degree is the modality inhibitance effect alleviated by the proposed model? 
\end{itemize}

In line with existing works~\cite{zhou2023tale, guo2024lgmrec}, we adopted three widely-used categories from the Amazon Review datasets~\cite{he2016vbpr} namely, (i) \textit{Baby}, (ii) \textit{Sports and Outdoors}, and (iii) \textit{Clothing, Shoes and Jewelry}, as released by McAuley et al., and the TikTok~\cite{wei2023multi} dataset for all our experiments. For brevity, we abbreviate the first three datasets as \textit{Baby}, \textit{Sports}, and \textit{Clothing}. In the above Amazon dataset, each item is associated with both visual and textual modality information, whereas the TikTok dataset additionally incorporates auditory features. As part of the pre-processing procedure, we adopted the standard five-core filtering strategy for both users and items, and report the statistics of the resulting dataset in Table~\ref{Raw_Dataset}.

In accordance with established practices~\cite{zhou2023tale, guo2024lgmrec}, we partitioned the interaction data into 80\%, 10\%, and 10\% for training, validation, and testing, respectively. In addition, to assess the retrieval quality in top-$K$ recommendation, we adopt the full-ranking evaluation protocol and utilized Recall@K and NDCG@K as the primary metrics for evaluation. The Recall metrics is defined as 
\begin{equation}
\operatorname{Recall@}K = 
\frac{1}{|\mathcal{U}_T|} 
\sum_{u \in \mathcal{U}_T}
\frac{ \sum_{i = 1}^{K} \mathds{1} \left\{ r_u(i) \in \mathcal{I}^u_t \right\} }
     {|\mathcal{I}^u_t|},
\end{equation}
where $r_u(i)$ indicates the position at which item $i$ is recommended to user $u$ within the ranked list, with $\mathds{1}(.)$ being the indicator function, returning a 1 if the condition holds true and 0 otherwise. The variable  $\mathcal{U}_T$ refers to the users belonging to the test set. NDCG, on the other hand, prioritizes the ranking of relevant items by discounting relevance scores based on item positions and is given by
\begin{equation}\label{eq:ndcg}
\mathrm{NDCG}@K \;=\;
\frac{1}{|\mathcal{U}_T|} 
\sum_{u \in \mathcal{U}_T}
\frac{\mathrm{DCG}@K(u)}{\mathrm{IDCG}@K(u)},
\end{equation}
where $\mathrm{IDCG}@K(u)$ denotes the ideal ranking of relevant items. Under this formulation, DCG@K takes the form of
\begin{equation}\label{eq:dcg}
\mathrm{DCG}@K(u) \;=\;
\sum_{i=1}^{K}
\frac{2^{\mathds{1}\left\{\,r_{u}(i)\in I^{u}_{t}\,\right\}}-1}
     {\log_{2}(i+1)},
\end{equation}
where $\log_{2}(\cdot)$ denotes the logarithmic discount term that reduces the impact of hits at lower positions.

\begin{table}[t]
\caption{Dataset Statistics}
\label{Raw_Dataset}
\centering
\begin{tabular}{lrrrr}
\toprule
Dataset & \#User & \#Item & \#Interaction & Density \\
\midrule
Baby & 19,445 & 7,050 & 160,792 & 0.117\% \\
Sports & 35,598 & 18,357 & 296,337 & 0.045\% \\
Clothing & 39,387 & 23,033 & 278,677 & 0.031\% \\
Tiktok & 9,308 & 6,710 & 68,722 & 0.110\% \\
\bottomrule
\end{tabular}
\end{table}
\begin{table*}[t]
\centering
\caption{Performance comparison of two general and ten multi-modal recommenders. Recall and NDCG are adopted as the primary evaluation metrics, with each abbreviated as R and N, respectively. To validate the consistency of the results, experiments were repeated with three different random seeds, and the observed improvements are statistically significant under a paired t-test with $p<0.01$.}
\label{tab:comparison}
\resizebox{\textwidth}{!}{
\begin{tabular}{l l l l l l l l l l l l l l l l l  l l l ll l l l l }
\toprule
\multirow{2}{*}{Datasets} & \multirow{2}{*}{Metrics} & \multicolumn{4}{c}{General Recommender} & \multicolumn{18}{c}{Multi-modal Recommender} \\ 
\cmidrule(lr){3-6} \cmidrule(lr){7-26}
& & \multicolumn{2}{c}{BPR} & \multicolumn{2}{c}{LightGCN} & \multicolumn{2}{c}{VBPR} & \multicolumn{2}{c}{LGCN-M} & \multicolumn{2}{c}{MMGCN} & \multicolumn{2}{c}{GRCN} & \multicolumn{2}{c}{SLMRec} & \multicolumn{2}{c}{BM3} & \multicolumn{2}{c}{FREEDOM} & \multicolumn{2}{c}{LGMRec} & \multicolumn{2}{c}{MIG-GT} & \multicolumn{2}{c}{MMWave} \\ \midrule
\multirow{6}{*}{Baby} & R@5  & \multicolumn{2}{c}{0.0235} & \multicolumn{2}{c}{0.0285} & \multicolumn{2}{c}{0.0266} & \multicolumn{2}{c}{0.0261} & \multicolumn{2}{c}{0.0261} & \multicolumn{2}{c}{0.0333} & \multicolumn{2}{c}{0.0341} & \multicolumn{2}{c}{0.0337} & \multicolumn{2}{c}{0.0377} & \multicolumn{2}{c}{\underline{0.0410}} & \multicolumn{2}{c}{0.0404} & \multicolumn{2}{c}{\textbf{0.0435}} \\ %
& R@10 & \multicolumn{2}{c}{0.0382} & \multicolumn{2}{c}{0.0453} & \multicolumn{2}{c}{0.0425} & \multicolumn{2}{c}{0.0428} & \multicolumn{2}{c}{0.0424} & \multicolumn{2}{c}{0.0534} & \multicolumn{2}{c}{0.0545} & \multicolumn{2}{c}{0.0548} & \multicolumn{2}{c}{0.0626} & \multicolumn{2}{c}{\underline{0.0653}} & \multicolumn{2}{c}{0.0650} & \multicolumn{2}{c}{\textbf{0.0690}} \\ %
& R@20 & \multicolumn{2}{c}{0.0595} & \multicolumn{2}{c}{0.0728} & \multicolumn{2}{c}{0.0663} & \multicolumn{2}{c}{0.0717} & \multicolumn{2}{c}{0.0668} & \multicolumn{2}{c}{0.0831} & \multicolumn{2}{c}{0.0837} & \multicolumn{2}{c}{0.0876} & \multicolumn{2}{c}{0.0986} & \multicolumn{2}{c}{\underline{0.1010}} & \multicolumn{2}{c}{0.1005} & \multicolumn{2}{c}{\textbf{0.1040}} \\ %
& N@5  & \multicolumn{2}{c}{0.0160} & \multicolumn{2}{c}{0.0191} & \multicolumn{2}{c}{0.0171} & \multicolumn{2}{c}{0.0170} & \multicolumn{2}{c}{0.0170} & \multicolumn{2}{c}{0.0223} & \multicolumn{2}{c}{0.0229} & \multicolumn{2}{c}{0.0227} & \multicolumn{2}{c}{0.0246} & \multicolumn{2}{c}{\underline{0.0273}} & \multicolumn{2}{c}{0.0271} & \multicolumn{2}{c}{\textbf{0.0290}} \\ %
& N@10 & \multicolumn{2}{c}{0.0207} & \multicolumn{2}{c}{0.0246} & \multicolumn{2}{c}{0.0223} & \multicolumn{2}{c}{0.0225} & \multicolumn{2}{c}{0.0223} & \multicolumn{2}{c}{0.0288} & \multicolumn{2}{c}{0.0296} & \multicolumn{2}{c}{0.0297} & \multicolumn{2}{c}{0.0327} & \multicolumn{2}{c}{\underline{0.0353}} & \multicolumn{2}{c}{0.0352} & \multicolumn{2}{c}{\textbf{0.0373}} \\ %
& N@20 & \multicolumn{2}{c}{0.0263} & \multicolumn{2}{c}{0.0317} & \multicolumn{2}{c}{0.0284} & \multicolumn{2}{c}{0.0300} & \multicolumn{2}{c}{0.0286} & \multicolumn{2}{c}{0.0365} & \multicolumn{2}{c}{0.0371} & \multicolumn{2}{c}{0.0381} & \multicolumn{2}{c}{0.0420} & \multicolumn{2}{c}{\underline{0.0444}} & \multicolumn{2}{c}{0.0443} & \multicolumn{2}{c}{\textbf{0.0463}} \\ \midrule %
\multirow{6}{*}{Sports} & R@5  & \multicolumn{2}{c}{0.0265} & \multicolumn{2}{c}{0.0350} & \multicolumn{2}{c}{0.0358} & \multicolumn{2}{c}{0.0338} & \multicolumn{2}{c}{0.0233} & \multicolumn{2}{c}{0.0382} & \multicolumn{2}{c}{0.0434} & \multicolumn{2}{c}{0.0395} & \multicolumn{2}{c}{0.0457} & \multicolumn{2}{c}{0.0462} & \multicolumn{2}{c}{\underline{0.0466}} & \multicolumn{2}{c}{\textbf{0.0499}} \\ %
& R@10 & \multicolumn{2}{c}{0.0417} & \multicolumn{2}{c}{0.0542} & \multicolumn{2}{c}{0.0561} & \multicolumn{2}{c}{0.0522} & \multicolumn{2}{c}{0.0386} & \multicolumn{2}{c}{0.0607} & \multicolumn{2}{c}{0.0676} & \multicolumn{2}{c}{0.0613} & \multicolumn{2}{c}{0.0724} & \multicolumn{2}{c}{0.0719} & \multicolumn{2}{c}{\underline{0.0738}} & \multicolumn{2}{c}{\textbf{0.0760}} \\ %
& R@20 & \multicolumn{2}{c}{0.0633} & \multicolumn{2}{c}{0.0837} & \multicolumn{2}{c}{0.0857} & \multicolumn{2}{c}{0.0791} & \multicolumn{2}{c}{0.0627} & \multicolumn{2}{c}{0.0922} & \multicolumn{2}{c}{0.1017} & \multicolumn{2}{c}{0.0940} & \multicolumn{2}{c}{0.1089} & \multicolumn{2}{c}{0.1107} & \multicolumn{2}{c}{\underline{0.1120}} & \multicolumn{2}{c}{\textbf{0.1148}} \\ %
& N@5  & \multicolumn{2}{c}{0.0183} & \multicolumn{2}{c}{0.0237} & \multicolumn{2}{c}{0.0241} & \multicolumn{2}{c}{0.0223} & \multicolumn{2}{c}{0.0154} & \multicolumn{2}{c}{0.0251} & \multicolumn{2}{c}{0.0294} & \multicolumn{2}{c}{0.0268} & \multicolumn{2}{c}{0.0303} & \multicolumn{2}{c}{0.0304} & \multicolumn{2}{c}{\underline{0.0313}} & \multicolumn{2}{c}{\textbf{0.0333}} \\ %
& N@10 & \multicolumn{2}{c}{0.0232} & \multicolumn{2}{c}{0.0300} & \multicolumn{2}{c}{0.0307} & \multicolumn{2}{c}{0.0284} & \multicolumn{2}{c}{0.0204} & \multicolumn{2}{c}{0.0325} & \multicolumn{2}{c}{0.0374} & \multicolumn{2}{c}{0.0339} & \multicolumn{2}{c}{0.0390} & \multicolumn{2}{c}{0.0388} & \multicolumn{2}{c}{\underline{0.0402}} & \multicolumn{2}{c}{\textbf{0.0419}} \\ %
& N@20 & \multicolumn{2}{c}{0.0288} & \multicolumn{2}{c}{0.0376} & \multicolumn{2}{c}{0.0384} & \multicolumn{2}{c}{0.0353} & \multicolumn{2}{c}{0.0266} & \multicolumn{2}{c}{0.0406} & \multicolumn{2}{c}{0.0462} & \multicolumn{2}{c}{0.0424} & \multicolumn{2}{c}{0.0484} & \multicolumn{2}{c}{0.0488} & \multicolumn{2}{c}{\underline{0.0501}} & \multicolumn{2}{c}{\textbf{0.0519}} \\ \midrule
\multirow{6}{*}{Clothing} & R@5 & \multicolumn{2}{c}{0.0127} & \multicolumn{2}{c}{0.0219} & \multicolumn{2}{c}{0.0191} & \multicolumn{2}{c}{0.0221} & \multicolumn{2}{c}{0.0137} & \multicolumn{2}{c}{0.0277} & \multicolumn{2}{c}{0.0291} & \multicolumn{2}{c}{0.0262} & \multicolumn{2}{c}{0.0396} & \multicolumn{2}{c}{0.0366} & \multicolumn{2}{c}{\underline{0.0410}} & \multicolumn{2}{c}{\textbf{0.0436}} \\
& R@10 & \multicolumn{2}{c}{0.0200} & \multicolumn{2}{c}{0.0338} & \multicolumn{2}{c}{0.0281} & \multicolumn{2}{c}{0.0357} & \multicolumn{2}{c}{0.0224} & \multicolumn{2}{c}{0.0428} & \multicolumn{2}{c}{0.0461} & \multicolumn{2}{c}{0.0418} & \multicolumn{2}{c}{\underline{0.0635}} & \multicolumn{2}{c}{0.0556} & \multicolumn{2}{c}{0.0629} & \multicolumn{2}{c}{\textbf{0.0673}} \\
& R@20 & \multicolumn{2}{c}{0.0295} & \multicolumn{2}{c}{0.0517} & \multicolumn{2}{c}{0.0410} & \multicolumn{2}{c}{0.0534} & \multicolumn{2}{c}{0.0362} & \multicolumn{2}{c}{0.0663} & \multicolumn{2}{c}{0.0696} & \multicolumn{2}{c}{0.0636} & \multicolumn{2}{c}{\underline{0.0938}} & \multicolumn{2}{c}{0.0843} & \multicolumn{2}{c}{0.0936} & \multicolumn{2}{c}{\textbf{0.0993}} \\
& N@5 & \multicolumn{2}{c}{0.0087} & \multicolumn{2}{c}{0.0146} & \multicolumn{2}{c}{0.0128} & \multicolumn{2}{c}{0.0145} & \multicolumn{2}{c}{0.0090} & \multicolumn{2}{c}{0.0178} & \multicolumn{2}{c}{0.0193} & \multicolumn{2}{c}{0.0174} & \multicolumn{2}{c}{0.0262} & \multicolumn{2}{c}{0.0243} & \multicolumn{2}{c}{\underline{0.0272}} & \multicolumn{2}{c}{\textbf{0.0288}} \\
& N@10 & \multicolumn{2}{c}{0.0111} & \multicolumn{2}{c}{0.0185} & \multicolumn{2}{c}{0.0157} & \multicolumn{2}{c}{0.0189} & \multicolumn{2}{c}{0.0118} & \multicolumn{2}{c}{0.0227} & \multicolumn{2}{c}{0.0249} & \multicolumn{2}{c}{0.0225} & \multicolumn{2}{c}{0.0340} & \multicolumn{2}{c}{0.0305} & \multicolumn{2}{c}{\underline{0.0343}} & \multicolumn{2}{c}{\textbf{0.0365}} \\
& N@20 & \multicolumn{2}{c}{0.0135} & \multicolumn{2}{c}{0.0230} & \multicolumn{2}{c}{0.0190} & \multicolumn{2}{c}{0.0233} & \multicolumn{2}{c}{0.0153} & \multicolumn{2}{c}{0.0287} & \multicolumn{2}{c}{0.0308} & \multicolumn{2}{c}{0.0281} & \multicolumn{2}{c}{0.0417} & \multicolumn{2}{c}{0.0377} & \multicolumn{2}{c}{\underline{0.0421}} & \multicolumn{2}{c}{\textbf{0.0446}} \\
\midrule
\multirow{6}{*}{Tiktok} & R@5 & \multicolumn{2}{c}{0.0184} & \multicolumn{2}{c}{0.0325} & \multicolumn{2}{c}{0.0151} & \multicolumn{2}{c}{0.0345} & \multicolumn{2}{c}{0.0309} & \multicolumn{2}{c}{0.0243} & \multicolumn{2}{c}{0.0394} & \multicolumn{2}{c}{\underline{0.0437}} & \multicolumn{2}{c}{0.0394} & \multicolumn{2}{c}{0.0305} & \multicolumn{2}{c}{0.0374} & \multicolumn{2}{c}{\textbf{0.0502}} \\
& R@10 & \multicolumn{2}{c}{0.0322} & \multicolumn{2}{c}{0.0529} & \multicolumn{2}{c}{0.0269} & \multicolumn{2}{c}{0.0558} & \multicolumn{2}{c}{0.0542} & \multicolumn{2}{c}{0.0374} & \multicolumn{2}{c}{0.0643} & \multicolumn{2}{c}{\underline{0.0722}} & \multicolumn{2}{c}{0.0571} & \multicolumn{2}{c}{0.0502}  & \multicolumn{2}{c}{0.0624} & \multicolumn{2}{c}{\textbf{0.0827}} \\
& R@20 & \multicolumn{2}{c}{0.0509} & \multicolumn{2}{c}{0.0785} & \multicolumn{2}{c}{0.0483} & \multicolumn{2}{c}{0.0886} & \multicolumn{2}{c}{0.0896} & \multicolumn{2}{c}{0.0693} & \multicolumn{2}{c}{0.1041} & \multicolumn{2}{c}{\underline{0.1067}} & \multicolumn{2}{c}{0.0991} & \multicolumn{2}{c}{0.0768} & \multicolumn{2}{c}{0.1005} & \multicolumn{2}{c}{\textbf{0.1188}} \\
& N@5 & \multicolumn{2}{c}{0.0114} & \multicolumn{2}{c}{0.0200} & \multicolumn{2}{c}{0.0094} & \multicolumn{2}{c}{0.0230} & \multicolumn{2}{c}{0.0152} & \multicolumn{2}{c}{0.0147} & \multicolumn{2}{c}{0.0241} & \multicolumn{2}{c}{\underline{0.0287}} & \multicolumn{2}{c}{0.0262} & \multicolumn{2}{c}{0.0204} & \multicolumn{2}{c}{0.0224} & \multicolumn{2}{c}{\textbf{0.0304}} \\
& N@10 & \multicolumn{2}{c}{0.0162} & \multicolumn{2}{c}{0.0266} & \multicolumn{2}{c}{0.0132} & \multicolumn{2}{c}{0.0298} & \multicolumn{2}{c}{0.0223} & \multicolumn{2}{c}{0.0188} & \multicolumn{2}{c}{0.0321} & \multicolumn{2}{c}{\underline{0.0379}} & \multicolumn{2}{c}{0.0319} & \multicolumn{2}{c}{0.0267} & \multicolumn{2}{c}{0.0303} & \multicolumn{2}{c}{\textbf{0.0408}} \\
& N@20 & \multicolumn{2}{c}{0.0207} & \multicolumn{2}{c}{0.0331} & \multicolumn{2}{c}{0.0186} & \multicolumn{2}{c}{0.0380} & \multicolumn{2}{c}{0.0312} & \multicolumn{2}{c}{0.0268} & \multicolumn{2}{c}{0.0420} & \multicolumn{2}{c}{\underline{0.0466}} & \multicolumn{2}{c}{0.0424} & \multicolumn{2}{c}{0.0335} & \multicolumn{2}{c}{0.0400} & \multicolumn{2}{c}{\textbf{0.0499}} \\
\bottomrule
\end{tabular}
}
\end{table*}
To assess the effectiveness of the proposed model, we benchmarked the proposed model against recent state-of-the-art approaches. These models can be classified into general and multi-modal recommenders, where the former rely largely on user-item interaction data for recommendation, while the latter utilize both user-item interactions and modality-enriched item information to enhance the quality of recommendations. In particular, the proposed model is benchmarked against the following general recommenders
\begin{itemize}
    \item BPR~\cite{rendle2012bpr}\textemdash a traditional collaborative filtering model that leverages pairwise matrix factorization (MF) for implicit interactions;
    \item LightGCN~\cite{he2020lightgcn}\textemdash a graph neural network (GNN)-based model that simplifies message propagation by omitting feature transformation and non-linear activation, thereby enhancing computational efficiency while ensuring effective recommendation performance,
\end{itemize}
and the following multi-modal recommenders
\begin{itemize}
    \item VBPR~\cite{he2016vbpr}\textemdash a model that enhances the MF framework by integrating visual features into the item encoding process;
    \item LGCN-M~\cite{wei2023multi}\textemdash  a modified LightGCN model~\cite{he2020lightgcn}, where modality features are injected into the graph convolutional layers for guiding the message propagation process;
    \item MMGCN~\cite{wei2019mmgcn}\textemdash a model that constructs a modality-specific graph to learn modal-specific representations, and subsequently applying late fusion to derive the final user and item representations;
    \item GRCN~\cite{wei2020graph}\textemdash a model that mitigates false-positive signals by introducing a graph refining layer which prunes noisy edges and refines the interaction graph structure before applying a graph convolutional network to learn high-order user and item representations;
    \item SLMRec~\cite{tao2022self}\textemdash a self-supervised GNN-based model designed to capture the underlying modality relations, thus enhancing the learned user and item representations;
    \item BM3~\cite{zhou2023bootstrap}\textemdash a model that employs a new self-supervised learning strategy seeking to minimize computational overhead and mitigate the impact of incorrect supervision signals arising from negative sampling; 
    \item FREEDOM~\cite{zhou2023tale}\textemdash a model that highlights redundancy in learning modality graph structures and freezes them prior to training, thereby reducing computational costs. It also employs a degree-aware edge pruning strategy to reduce noise from the interaction graph;
    \item LGMRec~\cite{guo2024lgmrec}\textemdash a recent model which leverages hypergraphs to capture user and item global dependencies. The generated global embeddings for users and items serve to complement and refine the embeddings derived from collaborative and modal-specific graphs.
    \item MIG-GT~\cite{hu2025modality}\textemdash a state-of-the-art multi-modal recommender, that employs a sampling-based global transformer to capture long-range dependencies for modality-specific graph learning, thereby improving node representations and retrieval performance.
\end{itemize}

In terms of implementation, we developed the proposed MMWave model and reproduced all baselines using the unified open-source MMRec framework~\cite{zhou2023mmrec}. For all selected baseline models, hyperparameter settings were tuned in accordance with the optimal configurations reported in the original publications. To ensure consistency in training and facilitate fair comparisons, we adopted the same random seed and set the embedding dimensionality of users and items to 64. All model parameters were initialized using Xavier~\cite{glorot2010understanding}, while the Adam optimizer was utilized for training. The entire training process spanned 1000 epochs with a batch size of 2048. To mitigate overfitting, early stoppage was triggered when no improvement was observed in the validation set for 20 consecutive epochs. Recall@20 was consistently adopted as the evaluation criterion for determining convergence across all models.

\subsection{Effectiveness of MMWave (RQ1)}
 Table~\ref{tab:comparison} highlights the performance achieved by two general, ten multi-modal baselines\footnote{At the time of paper acceptance, the complete implementation details for CMDL and R2MR were not yet publicly available to support a consistent and fair experimental comparison. MENTOR was successfully reproduced on Baby, while experiments on Sports and Clothing could not be conducted under the memory constraints of our NVIDIA GeForce RTX 2080 Ti GPU.}, and the proposed MMWave model. We note from these results that the proposed MMWave framework consistently outperforms all baseline models. Specifically, across the three Amazon datasets incorporating dual modalities (visual and textual) and the TikTok dataset encompassing three modalities (visual, textual and acoustic), MMWave achieves an average improvement of 6.48\% over all baseline models. The observed improvement arises from the extraction of essential modality features facilitated by the proposed wavelet modality mixer, coupled with the effective integration of high-order, structure-aware modality information through the multi-modal graph learning and adaptive modality regulator modules. This unified framework, in turn, accounts for the superior performance exhibited by MMWave.

Results presented also show that models that incorporate modality features tend to outperform those that rely solely on historical interactions. For instance, MRSs such as VBPR and LGCN-M, which extend BPR and LightGCN, respectively, tend to outperform their counterparts, implying that incorporating modality features is effective and essential for inferring user interests.

\begin{table}[t]
\caption{Performance comparison under modality ablation settings}
    \centering
    \begin{tabular}{llcccc}
        \toprule
        Datasets & Modality & R@10 & R@20 & N@10 & N@20 \\
        \midrule
        \multirow{4}{*}{Baby} 
        & Text & 0.0636 & 0.0991 & 0.0346 & 0.0436 \\
        & Visual & 0.0552 & 0.0872 & 0.0298 & 0.0381 \\
        & Fusion & 0.0578 & 0.0910 & 0.0314 & 0.0399 \\
        & All & \textbf{0.0690} & \textbf{0.1040} & \textbf{0.0373} & \textbf{0.0463} \\
        \midrule
        \multirow{4}{*}{Sports} 
        & Text & 0.0619 & 0.0937 & 0.0339 & 0.0421\\
        & Visual & 0.0712 & 0.1072 & 0.0382 & 0.0474 \\
        & Fusion & 0.0698 & 0.1057 & 0.0383 & 0.0476 \\
        & All & \textbf{0.0760} & \textbf{0.1148} & \textbf{0.0419} & \textbf{0.0519}\\
        \midrule
        \multirow{4}{*}{Clothing} 
        & Text & 0.0627 & 0.0935 & 0.0339 & 0.0417 \\
        & Visual & 0.0462 & 0.0698 & 0.0249 & 0.0308 \\
        & Fusion & 0.0549 & 0.0806 & 0.0302 & 0.0367\\
        & All
        & \textbf{0.0673} & \textbf{0.0993} & \textbf{0.0365} & \textbf{0.0446}\\
        \midrule
        \multirow{5}{*}{Tiktok} 
        & Text & 0.0781 & 0.1142 & 0.0378 & 0.0469 \\
        & Visual & 0.0739 & 0.1169 & 0.0379 & 0.0487 \\
        & Audio & 0.0706 & 0.1129 & 0.0375 & 0.0482\\
        & Fusion & 0.0676 & 0.1074 & 0.0362 & 0.0463\\
        & All & \textbf{0.0827} & \textbf{0.1188} & \textbf{0.0408} & \textbf{0.0499}\\
        \bottomrule
    \end{tabular}
    \label{tab:modality_ablation}
\end{table}
Notwithstanding the above, results also show that while graph-based fusion models demonstrate high performance in bi-modal scenarios, their effectiveness clearly reduces with the inclusion of more modalities. For the Amazon dataset, the recently proposed FREEDOM and LGMRec, which employ early and late fusion, respectively, outperform non-fusion models such as BM3 in a dual modality setting. Contrary to the Amazon dataset, BM3 demonstrates superior performance on the TikTok dataset, surpassing FREEDOM and LGMRec. This observation reinforces the earlier assertion regarding the persistent challenge in exploiting multiple modality features beyond two distinct modalities, which is attributed to the modality inhibitance effect. This is due to the shortcomings of existing fusion-based models to effectively leverage informative modality signals, thereby limiting their capacity to exploit modality features. Unlike existing approaches, MMWave distinguishes itself by emphasising the effective extraction and integration of modality features, which accounts for its competitive performance. 

\subsection{Ablation Studies}
To assess the contribution of each component, we segment the proposed MMWave into four distinct architectural variants: (i) MMWave, (ii) MMWave without wavelet modality mixer (w/o WMM), (iii) MMWave without structural-aware multi-modal graph learning (w/o SMGL), and (iv) MMWave without adaptive modality regulator (w/o AMR). It is worth noting that, in the final variant, the high-order modality features derived from~(\ref{n_item_mod}) and~(\ref{n_user_mod}) are directly averaged to obtain modality-side representations. 

\begin{figure}[t]
    \centering
    \includegraphics[width=\columnwidth]{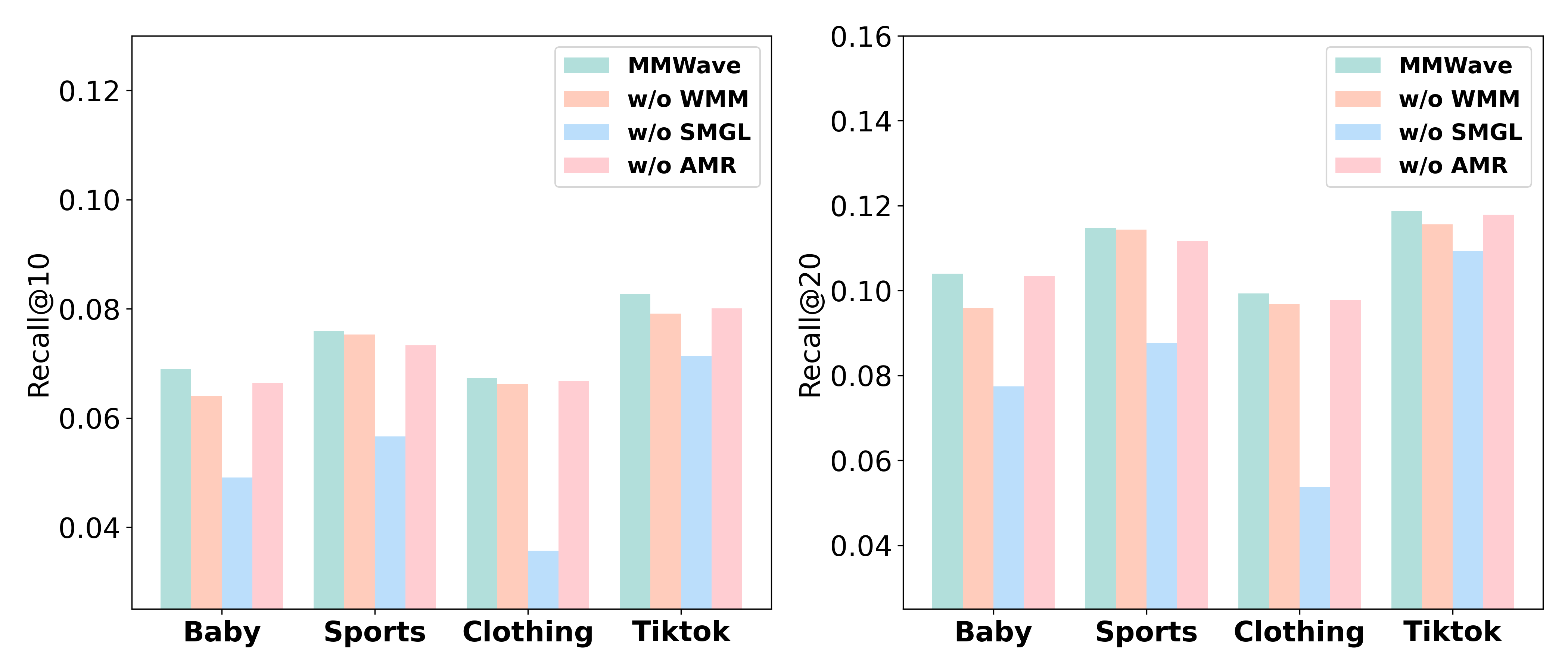}
    \caption{Ablation studies on the proposed MMWave.}
    \label{MMWave Ablation}
\end{figure}

As shown in Fig.~\ref{MMWave Ablation}, omitting any component leads to a consistent reduction in performance across all datasets. These empirical results demonstrate the critical role that each component plays within the proposed model. In particular, the wavelet modality filter aims to extract only the most informative signals pertaining to each modality, while the multi-modal graph learning module and the wavelet modality mixer are designed, respectively, to encode high-order structural features from uni-modal and fusion graphs, and to perform effective modality integration. As such, ablating any components will result in reduced performance.

The results also show that, among all key components, structure-aware learning is important to extract the latent structural signals underlying each modality. Among the four variants, the variant without the structure-aware multi-modal graph learning module results in the most detrimental impact on the model. This emphasizes the necessity of constructing both uni-modal and fused graphs for representation learning, and that structural information plays a key role in capturing the underlying associations between items. Investigating alternative fusion operations over uni-modal graphs represents a meaningful direction for future work to better understand their relative effectiveness in structure-aware learning.

In addition to the preceding observations, the need for robust feature extraction and integration becomes evidently more pronounced in multi-modal settings involving beyond two modalities. As depicted in the TikTok dataset, we observe that ablating either the wavelet modality mixer or the adaptive modality regulator, while retaining the other, results in similar substantial performance degradation, averaging a 7.6\% drop. This reflects the equally critical role of both components in extracting informative features from each modality, which becomes especially prominent when incorporating additional modalities.

To evaluate the individual contribution of each modality, we performed an ablation test in the form of a modality ablation study. More specifically, we expose the proposed model to a diverse set of uni-modal and fused input conditions comprising: text (sequential), visual (spatial), audio (sequential \& temporal), fused (complementary), and full encapsulating both the uni-modal and fused signals. With reference to Table~\ref{tab:modality_ablation}, the results indicate that excluding any modality results in performance degradation. It is clearly observed that the multi-modal (full) variant achieves the highest performance among all evaluated configurations. As each modality encapsulates distinct informative signals, removing any of them detrimentally impacts the performance of MMWave. In addition, we observe that even in the uni-modal version of our model, the results for TikTok and Sports datasets are comparable to those of recently proposed models such as BM3 and LGMRec. Consequently, this provides compelling evidence for MMWave's ability to effectively capitalize on each modality's features, thereby enhancing the overall performance.

\begin{figure}[t]
    \centering
    \includegraphics[width=\columnwidth]{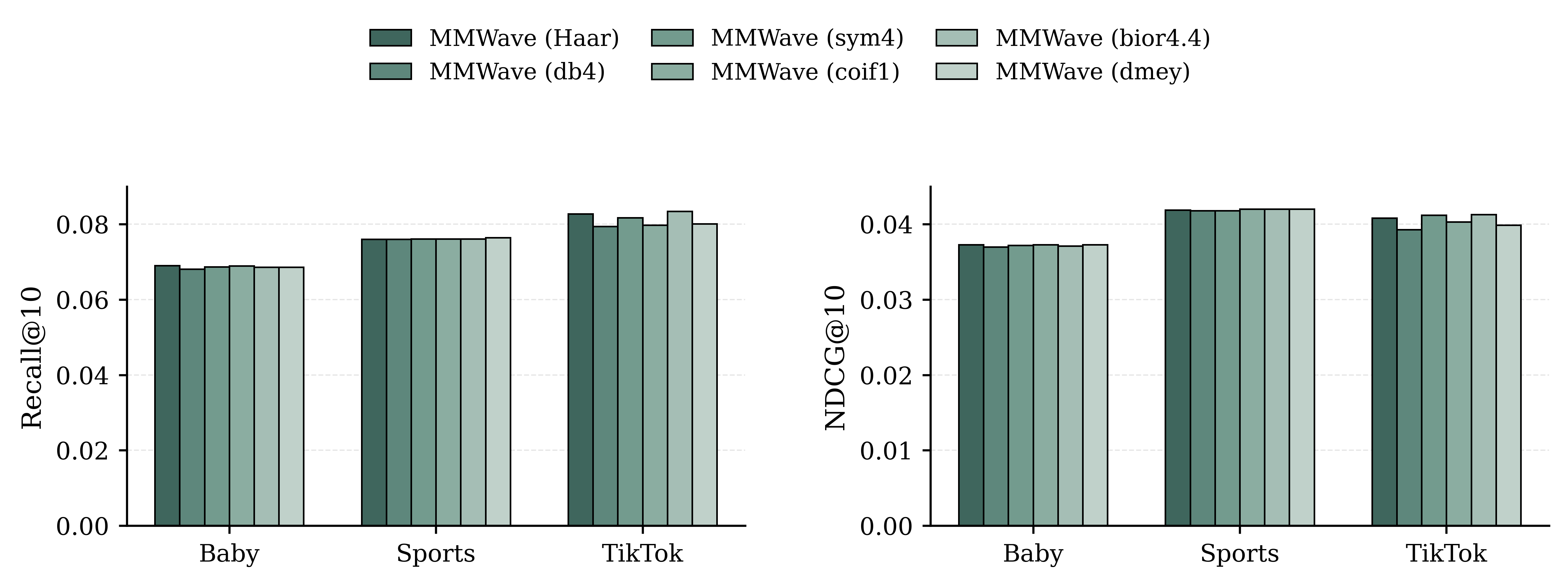}
    \caption{Comparison of MMWave with varying basis functions}     
    \label{MMWave_Basis_Function}
\end{figure}
\begin{figure}[t]
    \centering
    \includegraphics[width=\columnwidth]{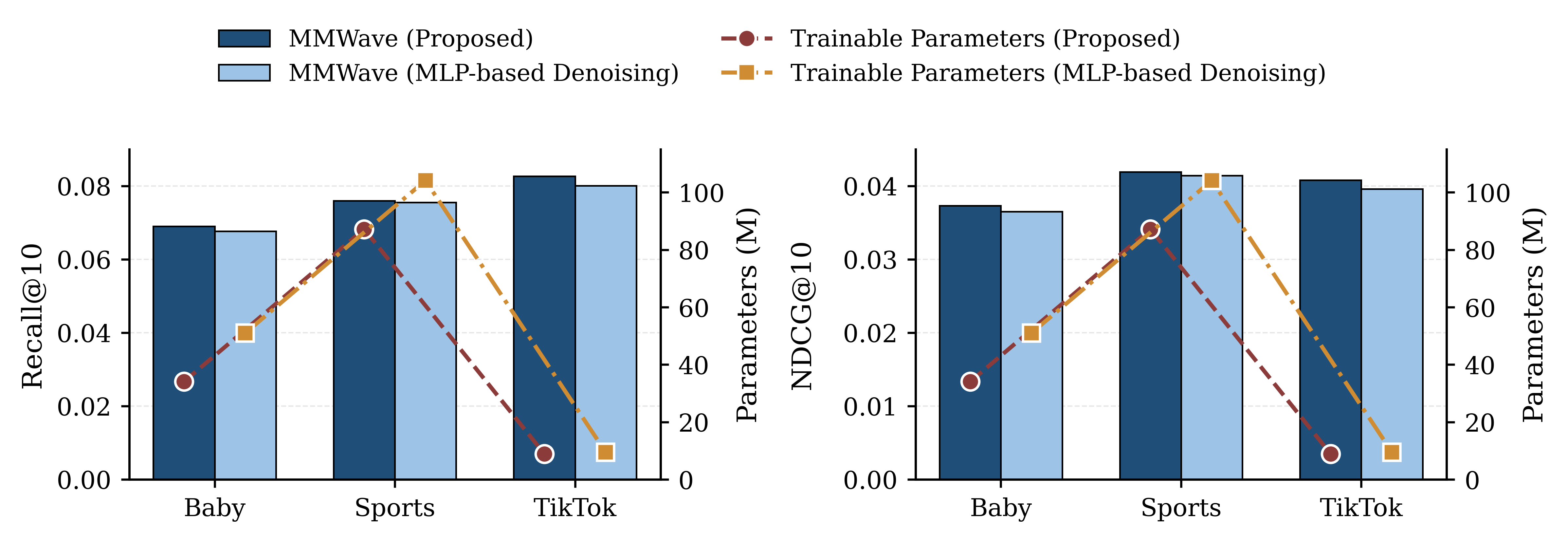}
    \caption{Comparison of MMWave with MLP-based Denoising}     
    \label{MMWave_Denoising_MLP}
\end{figure}
\begin{figure}[t]
    \centering
    \includegraphics[width=\columnwidth]{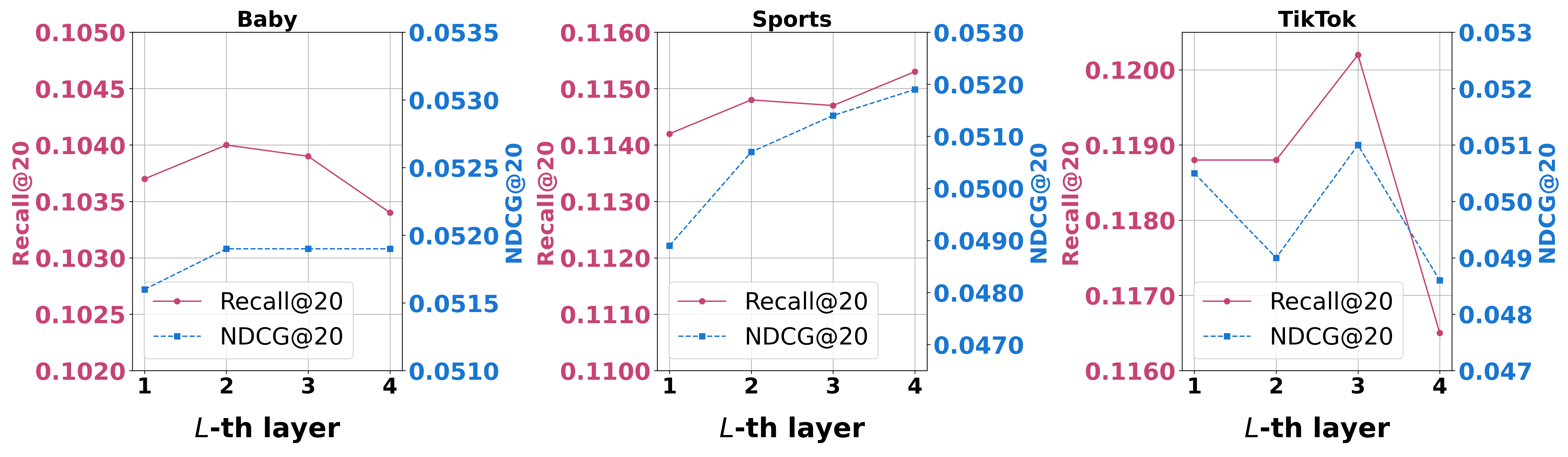}
    \caption{Variation of $L$ in MMWave.}
    \label{N_wavelet_layers}
\end{figure}
{
\begin{table*}[t]
\centering
\caption{Efficiency comparison of two general and five multi-modal recommenders.}
\label{tab:efficiency}
\resizebox{\textwidth}{!}{
\begin{tabular}{l l l l l l l l l l l l l l ll ll }
\toprule
\multirow{2}{*}{Datasets} & \multirow{2}{*}{Metrics} & \multicolumn{4}{c}{General Recommender} & \multicolumn{10}{c}{Multi-modal Recommender} \\ 
\cmidrule(lr){3-6} \cmidrule(lr){7-18}
& & \multicolumn{2}{c}{BPR} & \multicolumn{2}{c}{LightGCN} & \multicolumn{2}{c}{VBPR} & \multicolumn{2}{c}{SLMRec} & \multicolumn{2}{c}{BM3} & \multicolumn{2}{c}{FREEDOM} & \multicolumn{2}{c}{LGMRec} & \multicolumn{2}{c}{MMWave} \\ \midrule
\multirow{3}{*}{Baby} & Memory (GB)  & \multicolumn{2}{c}{1.24} & \multicolumn{2}{c}{1.32} & \multicolumn{2}{c}{1.42} & \multicolumn{2}{c}{1.63} & \multicolumn{2}{c}{1.64} & \multicolumn{2}{c}{1.66} & \multicolumn{2}{c}{1.58} & \multicolumn{2}{c}{2.55} \\ %
& Training (s/epoch) & \multicolumn{2}{c}{0.62} & \multicolumn{2}{c}{1.16} & \multicolumn{2}{c}{0.67} & \multicolumn{2}{c}{2.46} & \multicolumn{2}{c}{1.45} & \multicolumn{2}{c}{1.37} & \multicolumn{2}{c}{2.71} & \multicolumn{2}{c}{3.27} \\ %
& Evaluation (s/epoch) & \multicolumn{2}{c}{3.04} & \multicolumn{2}{c}{2.93} & \multicolumn{2}{c}{2.91} & \multicolumn{2}{c}{3.51} & \multicolumn{2}{c}{3.32} & \multicolumn{2}{c}{2.93} & \multicolumn{2}{c}{3.18} & \multicolumn{2}{c}{2.75}  \\ %
\midrule %
\multirow{3}{*}{Sports} & Memory (GB)  & \multicolumn{2}{c}{1.84} & \multicolumn{2}{c}{2.03} & \multicolumn{2}{c}{2.24} & \multicolumn{2}{c}{2.58} & \multicolumn{2}{c}{2.90} & \multicolumn{2}{c}{3.08} & \multicolumn{2}{c}{2.89} & \multicolumn{2}{c}{4.67}  \\ %
& Training (s/epoch) & \multicolumn{2}{c}{1.22} & \multicolumn{2}{c}{3.18} & \multicolumn{2}{c}{1.52} & \multicolumn{2}{c}{6.45} & \multicolumn{2}{c}{4.27} & \multicolumn{2}{c}{3.68} & \multicolumn{2}{c}{8.46} & \multicolumn{2}{c}{10.41}  \\ %
& Evaluation (s/epoch) & \multicolumn{2}{c}{5.33} & \multicolumn{2}{c}{5.05} & \multicolumn{2}{c}{5.65} & \multicolumn{2}{c}{5.29} & \multicolumn{2}{c}{5.85} & \multicolumn{2}{c}{4.84} & \multicolumn{2}{c}{5.66} & \multicolumn{2}{c}{5.60} \\ %
\midrule
\multirow{3}{*}{Tiktok} & Memory (GB) & \multicolumn{2}{c}{1.07} & \multicolumn{2}{c}{1.10} & \multicolumn{2}{c}{1.13} & \multicolumn{2}{c}{1.35} & \multicolumn{2}{c}{1.21} & \multicolumn{2}{c}{1.25} & \multicolumn{2}{c}{1.40} & \multicolumn{2}{c}{1.81}  \\
& Training (s/epoch) & \multicolumn{2}{c}{0.30} & \multicolumn{2}{c}{0.57} & \multicolumn{2}{c}{0.34} & \multicolumn{2}{c}{1.33} & \multicolumn{2}{c}{0.86} & \multicolumn{2}{c}{0.93} & \multicolumn{2}{c}{1.45} & \multicolumn{2}{c}{1.63} \\
& Evaluation (s/epoch) & \multicolumn{2}{c}{0.89} & \multicolumn{2}{c}{0.88} & \multicolumn{2}{c}{0.96} & \multicolumn{2}{c}{0.85} & \multicolumn{2}{c}{0.90} & \multicolumn{2}{c}{0.86} & \multicolumn{2}{c}{0.88} & \multicolumn{2}{c}{1.07} \\
\bottomrule
\end{tabular}
}
\end{table*}
}

With reference to the same table, we also note that the text modality exhibits strong discriminatory power when effectively exploited. In most cases within the e-commerce dataset, we observe that the exclusive use of text produces satisfactory results in comparison to other uni-modal variants. This is due to MMWave's ability to extract key features from detailed descriptions of the product, which accurately capture the semantic aspects of the items and enhance users' preference modeling. On the contrary, for short-video platforms such as Tiktok, we observe that the visual-based variant outperforms the text variant. This suggests that the visual modality plays a more significant role on video generation platforms, as users are primarily engaged through visual content while watching reels.

To investigate the impact of alternative wavelet basis functions, we replace the Haar basis adopted in our Wavelet Modality Mixer module with five alternatives: Db4, Sym4, Coif1, Bior4.4, and Dmey. In general, these wavelet bases differ in terms of their smoothness, symmetry, and orthogonality properties, resulting in the same signal being decomposed in different ways. As shown in Fig.~\ref{MMWave_Basis_Function}, for the Baby and Sports datasets, we observe that the choice of wavelet basis function generally does not lead to substantial variations in the performance of MMWave. For Tiktok, however, the variant using Bior4.4 wavelet basis function achieves higher performance than that of Haar. This may be attributed to the tri-modal features within Tiktok being more heterogeneous than those within the Amazon datasets and may contain more diverse semantic variations due to the dynamic nature of short-form videos. As such, the linear-phase property of the biorthogonal filters in Bior4.4 may be more effective in preserving subtle variations across image, text, and audio modalities. Overall, these results support the robustness of the proposed wavelet-based design, as the performance of MMWave remains relatively stable across different wavelet basis functions and is not highly sensitive to the choice of a specific basis.

To verify the efficacy of wavelet-based feature extraction relative to alternative denoising methods, we introduced a new MMWave variant that replaces the wavelet modality mixer with an MLP-based denoising module, denoted as MMWave (MLP-based Denoising). The module projects modality features into a lower-dimensional latent space through a linear transformation, followed by a non-linear activation function (e.g., LeakyReLU), and then maps them back to their original dimensionality. This design is inspired by the principle of denoising autoencoders, where noisy information is suppressed while informative modality signals are preserved. As shown in Fig.~\ref{MMWave_Denoising_MLP}, we observe that the proposed MMWave consistently outperforms the variant that adopts an MLP-based denoising mechanism across the Baby, Sports, and TikTok datasets. Despite an increase in the number of trainable parameters introduced by the MLP-based denoising variant, this modification does not yield any performance improvement. This implies that the effectiveness of MMWave is primarily attributable to its wavelet-based feature extraction mechanism, rather than to increased model capacity or regularization effect.
\begin{figure}[t]
    \centering
    \includegraphics[width=\columnwidth]{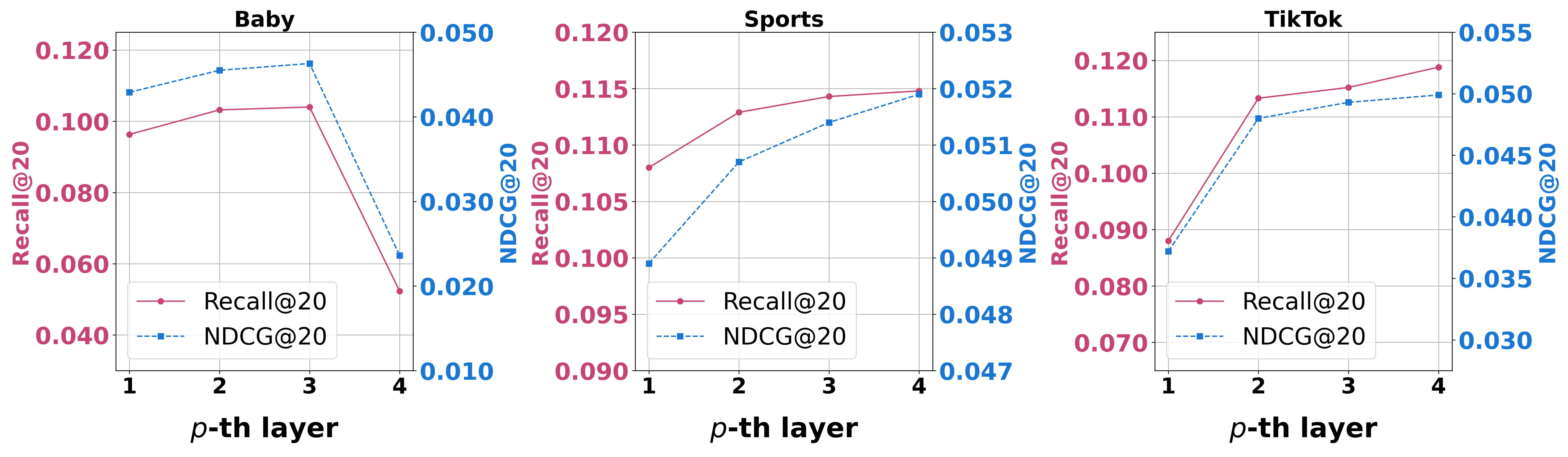}
    \caption{Variation of $p$ in MMWave.}
    \label{N_behavioral_layers}
\end{figure}

\begin{figure}[t]
    \centering
    \includegraphics[width=\columnwidth]{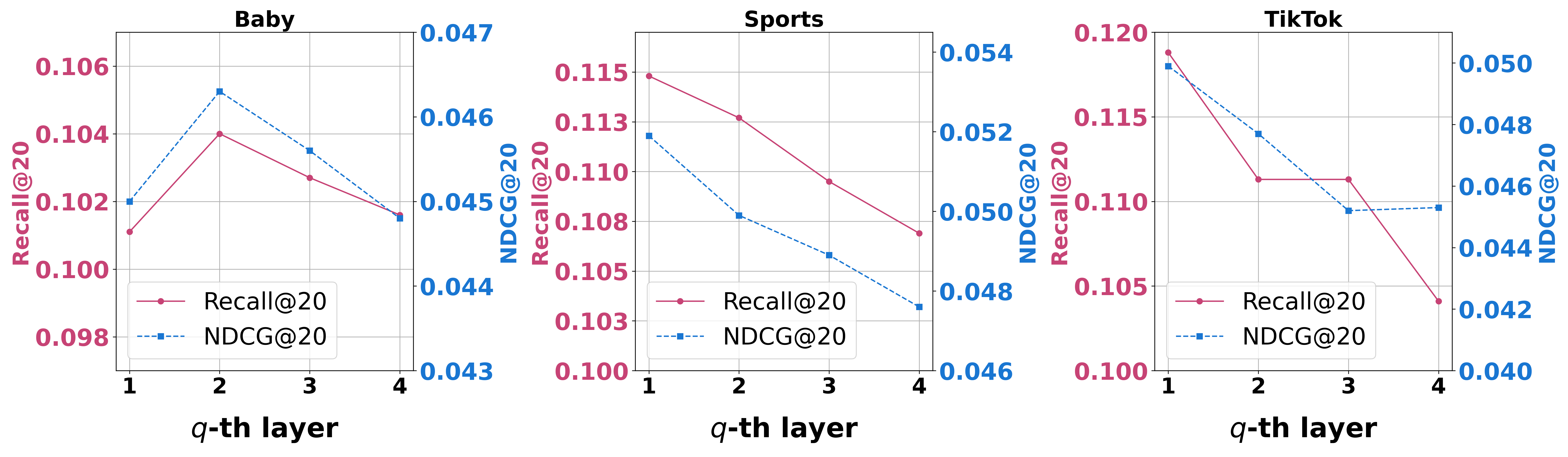}
    \caption{Variation of $q$ in MMWave.}
    \label{N_item_semantic_layers}
\end{figure}
\begin{figure}[t]
    \centering
    \includegraphics[width=\columnwidth]{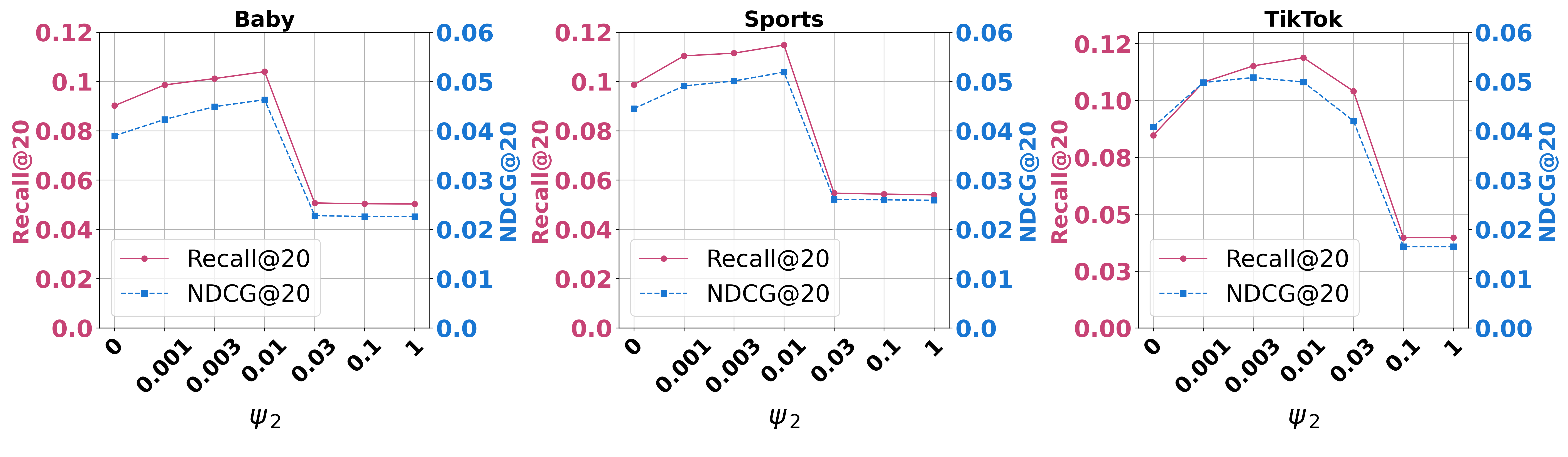}
    \caption{Variation of $\psi_{2}$ in MMWave.}
    \label{MMWave CL loss}
\end{figure}
\begin{figure}[t]
    \centering
    \includegraphics[scale=0.1750]{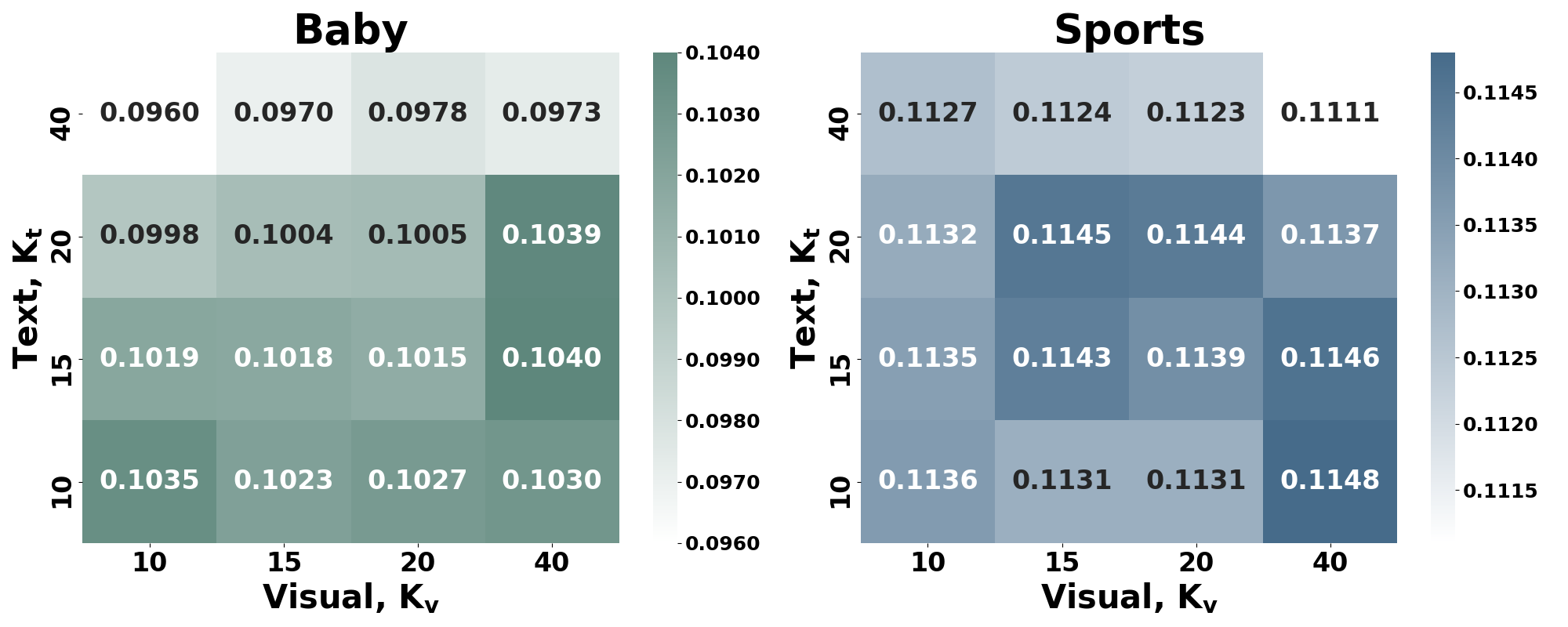}
    \caption{Variation of MMWave with $K_m$.}     
    \label{KNN_K_plots}
\end{figure}
\subsection{Hyperparameter Variations (RQ2)}
We first examine how the hyperparameter $L$, which determines the number of wavelet decomposition layers in Eq.~\ref{eq:1}, affects the overall performance of MMWave. To this end, we vary $L$ within the range $1\leq L \leq 4$. With reference to Fig.~\ref{eq:1}, these results indicate that a moderate level of multi-resolution decomposition (e.g., $2 \leq L \leq 3$) is generally sufficient in extracting key informative modality signals, whereas a lower level of decomposition (e.g., $L=1$) is insufficient to effectively discriminate noise and extract useful modality signals. We further observe that, on the TikTok dataset, performance degrades substantially under excessively deep decomposition (e.g., $L=4$), while the Sports dataset achieves the best performance at this setting. This observation suggests that the raw modality features in Sports exhibit stronger global semantic structure, whereas those in TikTok are more detail-sensitive and heterogeneous, likely due to the dynamic nature of reels. 

We next investigate the effect of varying the hyperparameter $p$ (which defines the number of GNN layers) with respect to the user-item behavioral graph. In particular, Fig.~\ref{N_behavioral_layers} shows the variation in performance with $p$. Across the Baby, Sports, and Tiktok dataset, we observe that increasing the number of GNN layers leads to better performance for the proposed model. This observation is consistent with prior findings, as using a single layer ($p=1$) captures only the immediate neighboring connections, whereas deeper layers ($p\geq 2$) capture long-range collaborative signals from high-order connectivities. However, for the Baby dataset, we observe a sharp decline in performance when increasing $p$ from $3$ to $4$, indicating the possibility of oversmoothing caused by the higher interaction density relative to other datasets. While most of the baseline methods~\cite{wei2019mmgcn, zhou2023bootstrap,zhou2023tale} achieve optimal performance at $p=2$, the optimal layer configurations for the three datasets in our model are found to be 3, 4, and 4, respectively. 

We next vary the $q$ hyperparameter within the range $1\leq q \leq 4$ as shown in Fig.~\ref{N_item_semantic_layers}. This hyperparameter defines the number of GNN layers used for item encoding in both the modality-specific and fusion graphs. From these results, we observe a consistent trend indicating that a shallow configuration of $q=1$ is sufficient to capture semantically correlated neighboring signals from the constructed uni-modal and structure-aware fusion graphs. However, as opposed to ~\cite{zhang2021mining,yu2023multi}, which reported that a light configuration is sufficient, MMWave achieves higher performance with a modestly deeper setting $q=2$ for the Baby dataset. This observation suggests that structural patterns may be more pertinent in Baby due to the more informative nature of its modality features (e.g., unstructured, complex textual annotations~\cite{zhu2024bringing}), thereby making structure-aware learning more critical.

To assess the influence of contrastive loss in conjunction with recommendation loss, we varied $0.0001\leq \psi_2 \leq 1$ as shown in Fig.~\ref{MMWave CL loss}. We observe that the absence of the contrastive loss when $\psi_2= 0$ leads to a noticeable drop in MMWave's performance. On the other hand, setting the contrastive weight to its upper limits of $\psi_2= 1$ results in the poorest performance. This performance trend implies that a balance is necessary since assigning too much weight to the contrastive loss may create undesirable imbalance and hinder the model's ability to optimize its primary recommendation objective effectively. In general, a small value ($\psi_2 = 0.01$) is sufficient to enhance the recommendation task.

To examine how variations in the graph construction threshold $K_m$ affect the uni-modal and fusion graphs, we tune the hyperparameter within the range $[10, 15, 20, 40]$ and results are presented in Fig.~\ref{KNN_K_plots}. Results from the Baby and Sports datasets show that a high visual ($K_v=40$) and moderately sized text ($K_t=10$ and 15) are sufficient to achieve optimal retrieval performance. However, as the top-k neighbors for text $K_t$ increase, the performance of MMWave reduces considerably. This performance reduction stems from unprocessed textual features that contain substantial redundancy, and a high $K_t$ would compromise and induce unnecessary high-order neighboring noise.

\begin{figure*}[t]
    \centering
    \includegraphics[width=0.70\textwidth]
    {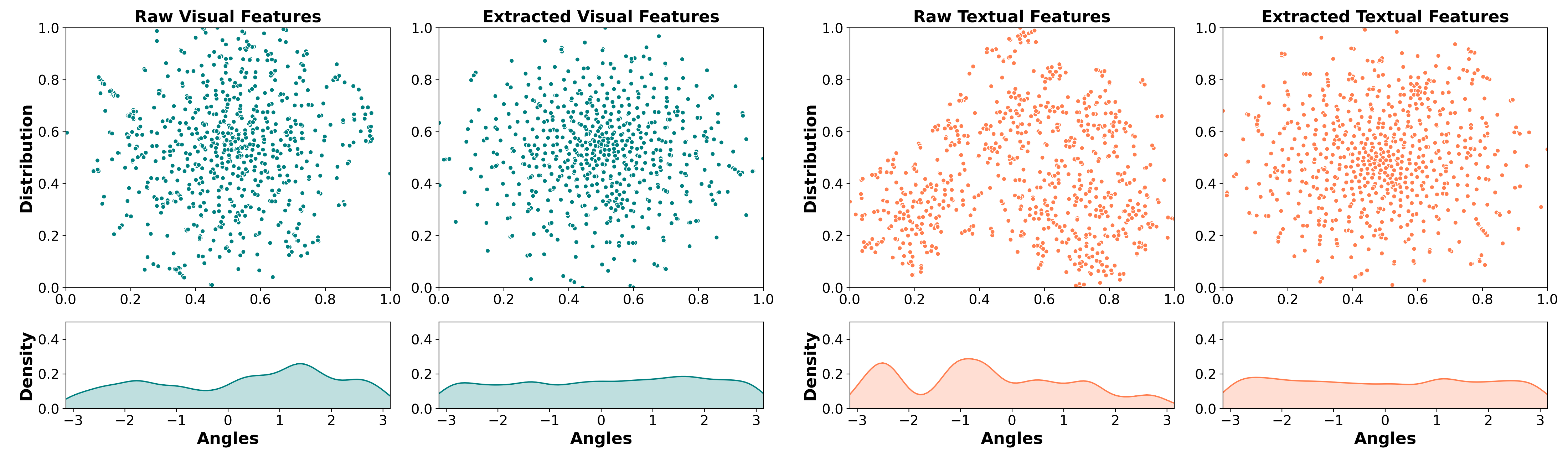}
    \caption{Distribution of Uni-modal features for Baby Dataset}
    \label{Baby_TSNE}
\end{figure*}
\begin{figure*}[t]
    \centering
    \includegraphics[width=1.0\textwidth]
    {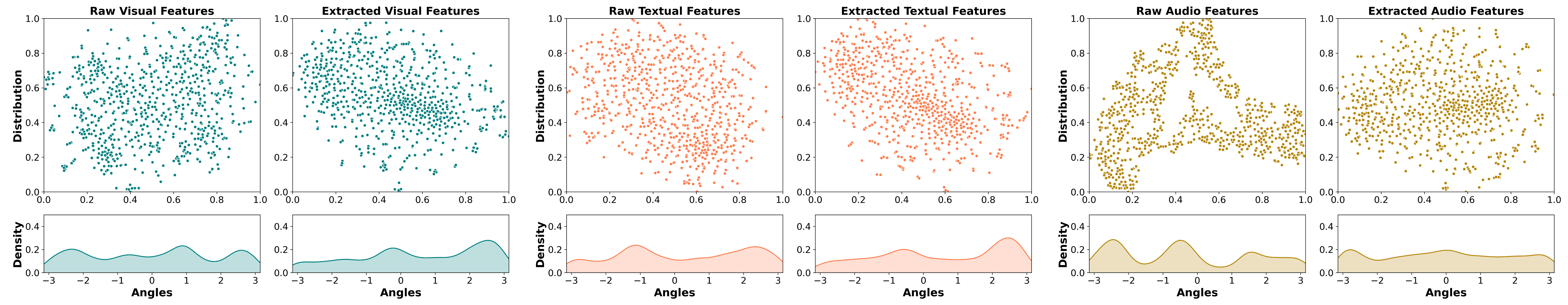}
    \caption{Distribution of Uni-modal features for Tiktok Dataset} 
    \label{Tiktok_TSNE}
\end{figure*}
\subsection{Practical Efficiency and Scalability (RQ4)}
To assess the practicality of MMWave for real-world industrial settings, we conduct an efficiency study based on three metrics: memory consumption, training time, and evaluation time, and benchmark it against seven other representative baselines. The reported training and evaluation times were averaged over the entire training and inference process and determined on an NVIDIA GeForce RTX 2080 Ti GPU. It is worth noting that runtime may vary across different GPU hardware, whereas the memory consumption of each model remains fixed under the same experimental configuration. In addition, this analysis excludes preprocessing time, including the eigenvalue decomposition required for fusion graph construction prior to training. With reference to Table~\ref{tab:efficiency}, and in terms of memory consumption and training/evaluation time, MMWave remains comparable to the existing state-of-the-art models, such as FREEDOM and LGMRec. Although the item–item multimodal views in MMWave are pre-computed during training similar to FREEDOM, the additional memory overhead arises primarily from the proposed adaptive modality regulator module. In particular, this module incorporates modality-specific and fusion MLPs to better capture the cross-modal relationships during modality integration, which plays an important role in mitigating one aspect of the Modality Inhibitance Effect. Overall, MMWave achieves improved performance with only modest computational overhead, indicating a favorable trade-off between effectiveness and scalability for practical real-world deployment.  

\begin{figure}
    \centering
    \includegraphics[width=0.85\columnwidth]
    {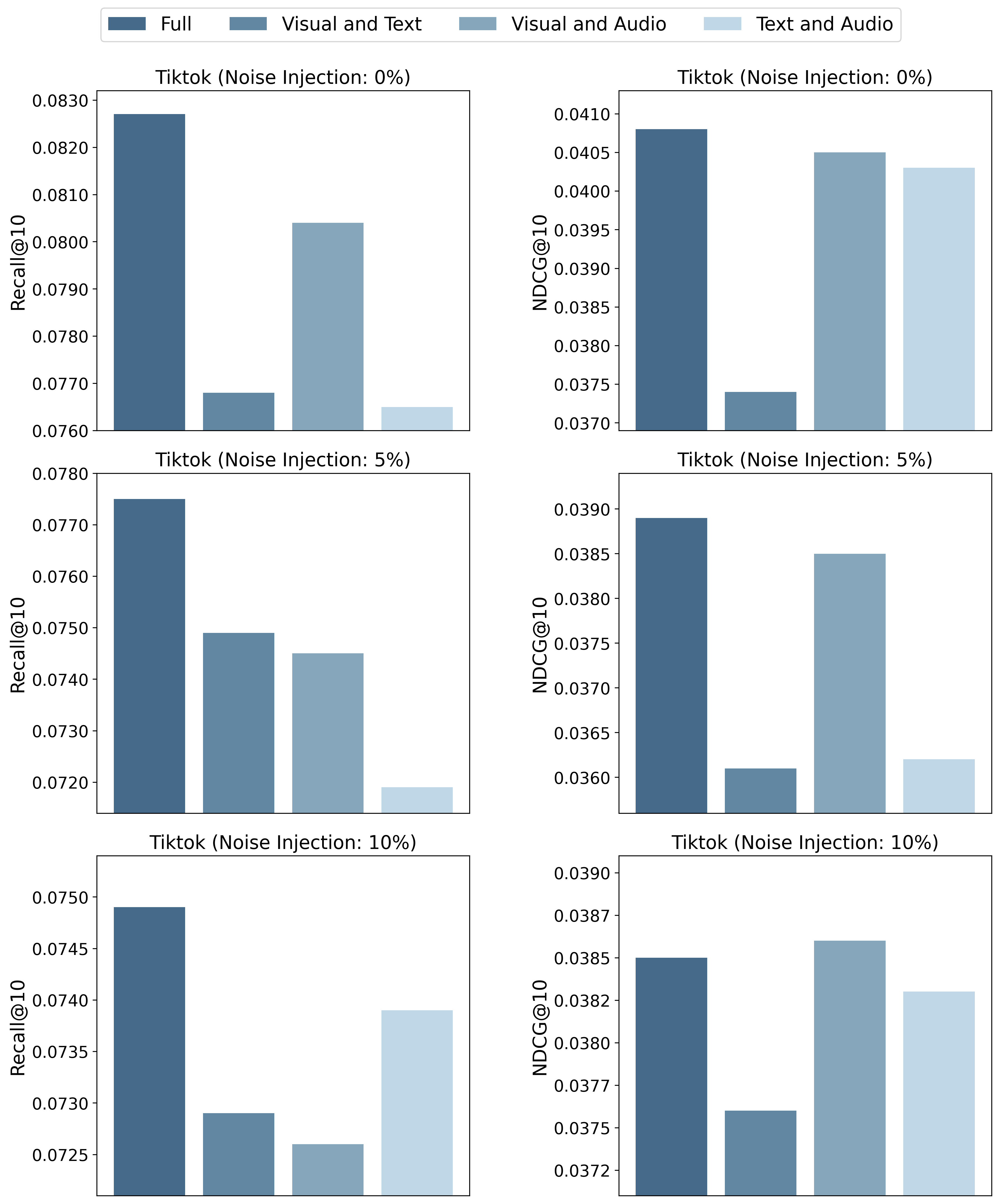}
    \caption{Performance variation across modality combinations under TikTok corruption settings}
    \label{fig:tiktok_corruption}
\end{figure}
\subsection{Impact of Feature Extraction and Integration on Modality Inhibition (RQ5)}
\label{final_subsection}
We investigate MMWave's ability to mitigate the modality inhibition effect from two critical aspects: (i) the quality of uni-modal features extracted from the wavelet modality mixer module and (ii) the observed performance shifts across distinct modality combinations.

To examine the representational differences qualitatively, we visualize and compare the modal-specific features generated by the wavelet modality mixer module against the unprocessed (raw) modality features. As shown in Figs.~(\ref{Baby_TSNE}) and~(\ref{Tiktok_TSNE}), we randomly sample a subset of 750 items from both the Baby and Tiktok datasets and subsequently employ t-SNE~\cite{van2008visualizing} to project the high-dimensional embeddings into a (lower) two-dimensional space. Using Gaussian kernel density estimation, we plotted the corresponding embedding distribution, with the estimation of $\arctan(y, x)$ over the unit circle $S^1$ shown at the bottom of each figure for better illustration.

As observed in the Baby and Tiktok datasets, we note that the wavelet extracted uni-modal features exhibit a more uniformly and equally spread distribution compared to the raw features. Notably, for Baby, the raw text features appear to be irregularly distributed, with a few sharp peaks observed in the density estimation plot. An even greater inconsistency is observed in the audio modality for the Tiktok dataset, where embeddings appear semantically fragmented, being both dispersed and densely concentrated across multiple regions. These observations collectively underscore the effectiveness of MMWave in extracting and attenuating essential uni-modal features, enabling each modality to exhibit its own unique semantic structure in the 2D space. 

To evaluate the MMWave's ability in modality integration, we conducted an experiment that centers on performance variations across different modality combinations. Considering that TikTok is the only dataset encompassing all three distinct modalities, we restrict the experiment to the TikTok dataset. To simulate varying degrees of multi-modal contamination, we follow existing practices~\cite{xv2024improving} and corrupt the Tiktok dataset with 0\%, 5\%, and 10\% multi-modal noise. Specifically, we randomly replaced item $i$ with an item $k$ ($i \ne k$) by substituting their raw modality features to mimic the feature corruption process. With different modality configurations, the resulting performance is depicted as shown in Fig.~\ref{fig:tiktok_corruption}. 

In contrast with the case study shown in Table~\ref{tab: case study}, MMWave demonstrates significantly higher capability to leverage all three modality features in clean and contaminated multi-modal scenarios\footnote{Further experiments on the MicroLens~\cite{ni2025content} dataset show consistent performance trends.}. This finding reinforces the effectiveness of our cross-weighting and modality integration mechanisms proposed in the adaptive modality regulator module of MMWave. It is also evident that, as the degree of contamination increases, the variant involving visual and audio performs comparably to the full modality-variant. This is due to visual and audio modalities learned and integrated by the proposed model being more discriminative in the context of short-video platforms, and that text features may be less reliable due to the spontaneous and arbitrary nature of user-generated captions. Identifying relevant types of textual descriptions for characterizing reels is, therefore, an important area for future research. Exploring simulated modalities also constitutes a promising research direction, as they may help address other important challenges in multi-modal recommendation, such as the missing-modality problem.

\section{Conclusion}
In this work, we described the modality inhibitance effect, which hinders the performance of existing fusion-based models in multi-modal recommendation systems. To address the identified issue, we propose the MMWave multi-modal recommender, which emphasizes two key aspects: the extraction of essential uni-modal features and the integration across semantically distinct modalities. In particular, we introduce an efficient yet effective wavelet-based approach to extract essential and informative modality signals for attenuation. To capture high-order behavioral and semantically affiliated modality signals, we then design a structure-aware multi-modal graph learning module that encapsulates structural patterns into modality representations. Finally, we propose a new adaptive modality regulator module that effectively integrates varying modalities through a designed cross-weighting (reciprocal) schema, enabling the model to capture complementary signals with higher precision. Extensive experiments conducted on four widely adopted datasets demonstrate the efficacy of MMWave, achieving an average improvement of 6.48\% across all metrics over state-of-the-art baselines. 

\bibliographystyle{IEEEtran}
\bibliography{references}

@article{khong2005,
  title={Selective-tap adaptive algorithms in the solution of the nonuniqueness problem for stereophonic acoustic echo cancellation},
  author={Khong, A. W. H.  and Naylor, P. A.},
  journal={IEEE Signal Process. Letters},
  volume={12},
number={4},
  pages={269--272},
  year={2005},
  publisher={IEEE}
}

@article{lin2012,
  title={A forced spectral diversity algorithm for speech dereverberation in the presence of near-common zeros},
  author={Lin, X. and Khong, A. W. H.  and Naylor, P. A.},
  journal={IEEE Trans. Audio, Speech, and Language Process.},
  volume={20},
  number={3},
  pages={888--899},
  year={2012}
}

@article{khong2006stereophonic,
  title={Stereophonic acoustic echo cancellation employing selective-tap adaptive algorithms},
  author={Khong, A. W. H. and Naylor, Patrick A},
  journal={IEEE Trans. Audio, Speech, and Language Process.},
  volume={14},
  number={3},
  pages={785--796},
  year={2006},
  publisher={IEEE}
}

@inproceedings{khong2008algorithms,
  title={Algorithms for identifying clusters of near-common zeros in multichannel blind system identification and equalization},
  author={Khong, A. W. H. and Lin, Xiang and Naylor, Patrick A},
  booktitle={2008 IEEE Trans. Acoustics, Speech and Signal Processing.},
  pages={389--392},
  year={2008},
  organization={IEEE}
}

@article{zhou2023comprehensive,
  title={A comprehensive survey on multimodal recommender systems: Taxonomy, evaluation, and future directions},
  author={Zhou, Hongyu and Zhou, Xin and Zeng, Zhiwei and Zhang, Lingzi and Miao, Chunyan},
  journal={IEEE Transactions on Pattern Analysis and Machine Intelligence},
  year={2026},
  volume={48},
  number={10},
  pages={13457--13476},
  publisher={IEEE}
}

@inproceedings{wang2019neural,
  title={Neural graph collaborative filtering},
  author={Wang, Xiang and He, Xiangnan and Wang, Meng and Feng, Fuli and Chua, Tat-Seng},
  booktitle={Proceedings of the 42nd international ACM SIGIR conference on Research and development in Information Retrieval},
  pages={165--174},
  year={2019}
}

@inproceedings{he2016vbpr,
  title={VBPR: visual bayesian personalized ranking from implicit feedback},
  author={He, Ruining and McAuley, Julian},
  booktitle={Proceedings of the AAAI Conference on Artificial Intelligence},
  volume={30},
  number={1},
  year={2016}
}

@inproceedings{chen2019personalized,
  title={Personalized fashion recommendation with visual explanations based on multimodal attention network: Towards visually explainable recommendation},
  author={Chen, Xu and Chen, Hanxiong and Xu, Hongteng and Zhang, Yongfeng and Cao, Yixin and Qin, Zheng and Zha, Hongyuan},
  booktitle={Proceedings of the 42nd International ACM SIGIR Conference on Research and Development in Information Retrieval},
  pages={765--774},
  year={2019}
}

@inproceedings{liu2019user,
  title={User diverse preference modeling by multimodal attentive metric learning},
  author={Liu, Fan and Cheng, Zhiyong and Sun, Changchang and Wang, Yinglong and Nie, Liqiang and Kankanhalli, Mohan},
  booktitle={Proceedings of the 27th ACM International Conference on Multimedia},
  pages={1526--1534},
  year={2019}
}

@inproceedings{zhang2021mining,
  title={Mining latent structures for multimedia recommendation},
  author={Zhang, Jinghao and Zhu, Yanqiao and Liu, Qiang and Wu, Shu and Wang, Shuhui and Wang, Liang},
  booktitle={Proceedings of the 29th ACM International Conference on Multimedia},
  pages={3872--3880},
  year={2021}
}

@article{zhang2022latent,
  title={Latent structure mining with contrastive modality fusion for multimedia recommendation},
  author={Zhang, Jinghao and Zhu, Yanqiao and Liu, Qiang and Zhang, Mengqi and Wu, Shu and Wang, Liang},
  journal={IEEE Transactions on Knowledge and Data Engineering},
  year={2022},
  publisher={IEEE}
}

@inproceedings{rendle2012bpr,
  title={{BPR}: Bayesian Personalized Ranking from Implicit Feedback},
  author={Rendle, Steffen and Freudenthaler, Christoph and Gantner, Zeno and Schmidt-Thieme, Lars},
  booktitle={Proceedings of the Twenty-Fifth Conference on Uncertainty in Artificial Intelligence},
  pages={452--461},
  year={2009},
  publisher={AUAI Press}
}

@inproceedings{he2020lightgcn,
  title={Lightgcn: Simplifying and powering graph convolution network for recommendation},
  author={He, Xiangnan and Deng, Kuan and Wang, Xiang and Li, Yan and Zhang, Yongdong and Wang, Meng},
  booktitle={Proceedings of the 43rd International ACM SIGIR Conference on Research and Development in Information Retrieval},
  pages={639--648},
  year={2020}
}

@inproceedings{wei2019mmgcn,
  title={MMGCN: Multi-modal graph convolution network for personalized recommendation of micro-video},
  author={Wei, Yinwei and Wang, Xiang and Nie, Liqiang and He, Xiangnan and Hong, Richang and Chua, Tat-Seng},
  booktitle={Proceedings of the 27th ACM International Conference on Multimedia},
  pages={1437--1445},
  year={2019}
}

@inproceedings{wei2020graph,
  title={Graph-refined convolutional network for multimedia recommendation with implicit feedback},
  author={Wei, Yinwei and Wang, Xiang and Nie, Liqiang and He, Xiangnan and Chua, Tat-Seng},
  booktitle={Proceedings of the 28th ACM International Conference on Multimedia},
  pages={3541--3549},
  year={2020}
}

@article{tao2022self,
  title={Self-supervised learning for multimedia recommendation},
  author={Tao, Zhulin and Liu, Xiaohao and Xia, Yewei and Wang, Xiang and Yang, Lifang and Huang, Xianglin and Chua, Tat-Seng},
  journal={IEEE Transactions on Multimedia},
  year={2022},
  publisher={IEEE}
}

@inproceedings{zhou2023bootstrap,
  title={Bootstrap latent representations for multi-modal recommendation},
  author={Zhou, Xin and Zhou, Hongyu and Liu, Yong and Zeng, Zhiwei and Miao, Chunyan and Wang, Pengwei and You, Yuan and Jiang, Feijun},
  booktitle={Proceedings of the ACM Web Conference 2023},
  pages={845--854},
  year={2023}
}

@inproceedings{zhou2023tale,
  title={A tale of two graphs: Freezing and denoising graph structures for multimodal recommendation},
  author={Zhou, Xin and Shen, Zhiqi},
  booktitle={Proceedings of the 31st ACM International Conference on Multimedia},
  pages={935--943},
  year={2023}
}

@inproceedings{yu2023multi,
  title={Multi-view graph convolutional network for multimedia recommendation},
  author={Yu, Penghang and Tan, Zhiyi and Lu, Guanming and Bao, Bing-Kun},
  booktitle={Proceedings of the 31st ACM International Conference on Multimedia},
  pages={6576--6585},
  year={2023}
}

@inproceedings{zhou2023enhancing,
  title={Enhancing dyadic relations with homogeneous graphs for multimodal recommendation},
  author={Zhou, Hongyu and Zhou, Xin and Zhang, Lingzi and Shen, Zhiqi},
  booktitle={ECAI 2023: 26th European Conference on Artificial Intelligence},
  pages={3123--3130},
  year={2023},
  organization={SAGE Publications 1 Oliver's Yard, 55 City Road, London, EC1Y 1SP}
}

@inproceedings{reimers2019sentence,
  title={Sentence-bert: Sentence embeddings using siamese bert-networks},
  author={Reimers, Nils and Gurevych, Iryna},
  booktitle={Proceedings of the 2019 Conference on Empirical Methods in Natural Language Processing and the 9th International Joint Conference on Natural Language Processing (EMNLP-IJCNLP)},
  pages = "3982--3992",
  year={2019},
  publisher={Association for Computational Linguistics},
  doi = "10.18653/v1/D19-1410",
}

@inproceedings{simonyan2014very,
  title={Very deep convolutional networks for large-scale image recognition},
  author={Simonyan, Karen and Zisserman, Andrew},
  booktitle={International Conference on Learning Representations},
  year={2015}
}

@inproceedings{liu2017deepstyle,
  title={Deepstyle: Learning user preferences for visual recommendation},
  author={Liu, Qiang and Wu, Shu and Wang, Liang},
  booktitle={{Proceedings of the 40th International ACM SIGIR Conference on Research and Development in Information Retrieval}},
  pages={841--844},
  year={2017}
}

@inproceedings{kipf2016semi,
  title={Semi-Supervised Classification with Graph Convolutional Networks},
  author={Kipf, Thomas N. and Welling, Max},
  booktitle={International Conference on Learning Representations},
  year={2017}
}

@article{van2008visualizing,
  title={Visualizing data using t-SNE.},
  author={Van der Maaten, Laurens and Hinton, Geoffrey},
  journal={Journal of Machine Learning Research},
  volume={9},
  number={11},
  year={2008}
}

@inproceedings{zhou2023mmrec,
  title={Mmrec: Simplifying multimodal recommendation},
  author={Zhou, Xin},
  booktitle={Proceedings of the 5th ACM International Conference on Multimedia in Asia Workshops},
  pages={1--2},
  year={2023}
}

@article{oord2018representation,
  title={Representation learning with contrastive predictive coding},
  author={Oord, Aaron van den and Li, Yazhe and Vinyals, Oriol},
  journal={arXiv preprint arXiv:1807.03748},
  year={2018}
}

@inproceedings{glorot2010understanding,
  title={Understanding the difficulty of training deep feedforward neural networks},
  author={Glorot, Xavier and Bengio, Yoshua},
  booktitle={Proceedings of the Thirteenth International Conference on Artificial Intelligence and Statistics},
  pages={249--256},
  year={2010},
  organization={JMLR Workshop and Conference Proceedings}
}

@article{zhang2024multimodal,
  title={Multimodal fusion on low-quality data: A comprehensive survey},
  author={Zhang, Qingyang and Wei, Yake and Han, Zongbo and Fu, Huazhu and Peng, Xi and Hu, Qinghua and Deng, Cheng and Xu, Cai and Wen, Jie and Hu, Di and others},
  journal={Information Fusion},
  pages={104437},
  year={2026},
  publisher={Elsevier}
}

@article{chen2009fast,
  title={Fast Approximate kNN Graph Construction for High Dimensional Data via Recursive Lanczos Bisection.},
  author={Chen, Jie and Fang, Haw-ren and Saad, Yousef},
  journal={Journal of Machine Learning Research},
  volume={10},
  number={9},
  year={2009}
}

@inproceedings{ma2024leveraging,
  title={Leveraging multimodal features and item-level user feedback for bundle construction},
  author={Ma, Yunshan and Liu, Xiaohao and Wei, Yinwei and Tao, Zhulin and Wang, Xiang and Chua, Tat-Seng},
  booktitle={Proceedings of the 17th ACM International Conference on Web Search and Data Mining},
  pages={510--519},
  year={2024}
}

@inproceedings{dai2022towards,
  title={Towards robust graph neural networks for noisy graphs with sparse labels},
  author={Dai, Enyan and Jin, Wei and Liu, Hui and Wang, Suhang},
  booktitle={Proceedings of the Fifteenth ACM International Conference on Web Search and Data Mining},
  pages={181--191},
  year={2022}
}

@inproceedings{zhao2022joint,
  title={Joint learning of e-commerce search and recommendation with a unified graph neural network},
  author={Zhao, Kai and Zheng, Yukun and Zhuang, Tao and Li, Xiang and Zeng, Xiaoyi},
  booktitle={Proceedings of the Fifteenth ACM International Conference on Web Search and Data Mining},
  pages={1461--1469},
  year={2022}
}

@inproceedings{mao2021ultragcn,
  title={UltraGCN: ultra simplification of graph convolutional networks for recommendation},
  author={Mao, Kelong and Zhu, Jieming and Xiao, Xi and Lu, Biao and Wang, Zhaowei and He, Xiuqiang},
  booktitle={Proceedings of the 30th ACM International Conference on Information \& Knowledge Management},
  pages={1253--1262},
  year={2021}
}

@inproceedings{wei2023multi,
  title={Multi-modal self-supervised learning for recommendation},
  author={Wei, Wei and Huang, Chao and Xia, Lianghao and Zhang, Chuxu},
  booktitle={Proceedings of the ACM Web Conference 2023},
  pages={790--800},
  year={2023}
}

@inproceedings{guo2024lgmrec,
  title={LGMRec: Local and Global Graph Learning for Multimodal Recommendation},
  author={Guo, Zhiqiang and Li, Jianjun and Li, Guohui and Wang, Chaoyang and Shi, Si and Ruan, Bin},
  booktitle={Proceedings of the AAAI Conference on Artificial Intelligence},
  volume={38},
  number={8},
  pages={8454--8462},
  year={2024}
}

@inproceedings{caisimple,
  title={LightGCL: Simple Yet Effective Graph Contrastive Learning for Recommendation},
  author={Cai, Xuheng and Huang, Chao and Xia, Lianghao and Ren, Xubin},
  booktitle={The Eleventh International Conference on Learning Representations},
  year={2023}
}

@inproceedings{zhang2024modality,
  title={Modality-Balanced Learning for Multimedia Recommendation},
  author={Zhang, Jinghao and Liu, Guofan and Liu, Qiang and Wu, Shu and Wang, Liang},
  booktitle={Proceedings of the 32nd ACM International Conference on Multimedia},
  pages={7551--7560},
  year={2024}
}

@incollection{zhang2019wavelet,
  title={Wavelet transform},
  author={Zhang, Dengsheng},
  booktitle={Fundamentals of image data mining: Analysis, Features, Classification and Retrieval},
  pages={35--44},
  year={2019},
  publisher={Springer}
}

@inproceedings{oka2025wavelet,
  title={Wavelet-based positional representation for long context},
  author={Oka, Yui and Hasegawa, Taku and Nishida, Kyosuke and Saito, Kuniko},
  booktitle={International Conference on Learning Representations},
  volume={2025},
  pages={77657--77684},
  year={2025}
}

@inproceedings{salama2024semantic,
  title={Semantic Compression for Word and Sentence Embeddings using Discrete Wavelet Transform},
  author={Salama, Rana and Youssef, Abdou and Diab, Mona},
  booktitle={Findings of the Association for Computational Linguistics ACL 2024},
  pages={15963--15977},
  year={2024}
}

@article{mallat1989theory,
  title={A theory for multiresolution signal decomposition: the wavelet representation},
  author={Mallat, Stephane G},
  journal={IEEE transactions on pattern analysis and machine intelligence},
  volume={11},
  number={7},
  pages={674--693},
  year={1989},
  publisher={Ieee}
}

@inproceedings{li2020wavelet,
  title={Wavelet integrated CNNs for noise-robust image classification},
  author={Li, Qiufu and Shen, Linlin and Guo, Sheng and Lai, Zhihui},
  booktitle={Proceedings of the IEEE/CVF conference on computer vision and pattern recognition},
  pages={7245--7254},
  year={2020}
}

@inproceedings{al2021towards,
  title={Towards a better understanding of noise in natural language processing},
  author={Al Sharou, Khetam and Li, Zhenhao and Specia, Lucia},
  booktitle={Proceedings of the international conference on recent advances in natural language processing (RANLP 2021)},
  pages={53--62},
  year={2021}
}

@article{polspoel2022importance,
  title={The importance of extended high-frequency speech information in the recognition of digits, words, and sentences in quiet and noise},
  author={Polspoel, Sigrid and Kramer, Sophia E and van Dijk, Bas and Smits, Cas},
  journal={Ear and hearing},
  volume={43},
  number={3},
  pages={913--920},
  year={2022},
  publisher={LWW}
}

@inproceedings{ong2025spectrum,
  title={Spectrum-based modality representation fusion graph convolutional network for multimodal recommendation},
  author={Ong, Rongqing Kenneth and Khong, Andy W. H.},
  booktitle={Proceedings of the Eighteenth ACM International Conference on Web Search and Data Mining},
  pages={773--781},
  year={2025}
}

@article{xu2025survey,
  title={A survey on multimodal recommender systems: Recent advances and future directions},
  author={Xu, Jinfeng and Chen, Zheyu and Yang, Shuo and Li, Jinze and Wang, Wei and Hu, Xiping and Hoi, Steven and Ngai, Edith},
  journal={IEEE Transactions on Multimedia},
  year={2026},
  volume={28},
  pages={7189--7203},
  publisher={IEEE}
}

@inproceedings{zhu2024bringing,
  title={Bringing multimodality to Amazon visual search system},
  author={Zhu, Xinliang and Huang, Sheng-Wei and Ding, Han and Yang, Jinyu and Chen, Kelvin and Zhou, Tao and Neiman, Tal and Xie, Ouye and Tran, Son and Yao, Benjamin and others},
  booktitle={Proceedings of the 30th ACM SIGKDD Conference on Knowledge Discovery and Data Mining},
  pages={6390--6399},
  year={2024}
}

@inproceedings{xv2024improving,
  title={Improving Multi-modal Recommender Systems by Denoising and Aligning Multi-modal Content and User Feedback},
  author={Xv, Guipeng and Li, Xinyu and Xie, Ruobing and Lin, Chen and Liu, Chong and Xia, Feng and Kang, Zhanhui and Lin, Leyu},
  booktitle={Proceedings of the 30th ACM SIGKDD Conference on Knowledge Discovery and Data Mining},
  pages={3645--3656},
  year={2024}
}

@inproceedings{peng2022balanced,
  title={Balanced multimodal learning via on-the-fly gradient modulation},
  author={Peng, Xiaokang and Wei, Yake and Deng, Andong and Wang, Dong and Hu, Di},
  booktitle={Proceedings of the IEEE/CVF conference on computer vision and pattern recognition},
  pages={8238--8247},
  year={2022}
}

@article{liang2022mind,
  title={Mind the gap: Understanding the modality gap in multi-modal contrastive representation learning},
  author={Liang, Victor Weixin and Zhang, Yuhui and Kwon, Yongchan and Yeung, Serena and Zou, James Y},
  journal={Advances in Neural Information Processing Systems},
  volume={35},
  pages={17612--17625},
  year={2022}
}

@article{li2024multimodal,
  title={Multimodal graph causal embedding for multimedia-based recommendation},
  author={Li, Shuaiyang and Xue, Feng and Liu, Kang and Guo, Dan and Hong, Richang},
  journal={IEEE Transactions on Knowledge and Data Engineering},
  volume={36},
  number={12},
  pages={8842--8858},
  year={2024},
  publisher={IEEE}
}

@article{zhang2025survey,
  title={A Survey on Side Information-driven Session-based Recommendation: From a Data-centric Perspective},
  author={Zhang, Xiaokun and Xu, Bo and Li, Chenliang and He, Bowei and Lin, Hongfei and Ma, Chen and Ma, Fenglong},
  journal={IEEE Transactions on Knowledge and Data Engineering},
  year={2025},
  publisher={IEEE}
}

@article{zhang2023beyond,
  title={Beyond co-occurrence: Multi-modal session-based recommendation},
  author={Zhang, Xiaokun and Xu, Bo and Ma, Fenglong and Li, Chenliang and Yang, Liang and Lin, Hongfei},
  journal={IEEE Transactions on Knowledge and Data Engineering},
  volume={36},
  number={4},
  pages={1450--1462},
  year={2023},
  publisher={IEEE}
}

@article{lin2025contrastive,
  title={Contrastive modality-disentangled learning for multimodal recommendation},
  author={Lin, Xixun and Liu, Rui and Cao, Yanan and Zou, Lixin and Li, Qian and Wu, Yongxuan and Liu, Yang and Yin, Dawei and Xu, Guandong},
  journal={ACM Trans. Information Systems},
  volume={43},
  number={3},
  pages={1--31},
  year={2025},
  publisher={ACM New York, NY}
}

@inproceedings{tang2025r2mr,
  title={R2MR: Review and Rewrite Modality for Recommendation},
  author={Tang, Gu and Wang, Jinghe and Gan, Xiaoying and Lu, Bin and Zhao, Ze and Fu, Luoyi and Wang, Xinbing and Zhou, Chenghu},
  booktitle={Proc. 31st ACM SIGKDD Conf. on Knowledge Discovery and Data Mining V. 1},
  pages={1337--1348},
  year={2025}
}

@inproceedings{xu2025mentor,
  title={Mentor: multi-level self-supervised learning for multimodal recommendation},
  author={Xu, Jinfeng and Chen, Zheyu and Yang, Shuo and Li, Jinze and Wang, Hewei and Ngai, Edith CH},
  booktitle={Proc. AAAI Conf. Artificial Intelligence},
  volume={39},
  number={12},
  pages={12908--12917},
  year={2025}
}

@inproceedings{hu2025modality,
  title={Modality-independent graph neural networks with global transformers for multimodal recommendation},
  author={Hu, Jun and Hooi, Bryan and He, Bingsheng and Wei, Yinwei},
  booktitle={Proc. AAAI Conf. Artificial Intelligence},
  volume={39},
  number={11},
  pages={11790--11798},
  year={2025}
}

@inproceedings{pomo2025recommender,
  title={Do recommender systems really leverage multimodal content? A comprehensive analysis on multimodal representations for recommendation},
  author={Pomo, Claudio and Attimonelli, Matteo and Danese, Danilo and Narducci, Fedelucio and Di Noia, Tommaso},
  booktitle={Proc. 34th ACM Int’l Conf. Information and Knowledge Management},
  pages={2377--2387},
  year={2025}
}

@article{zhou2025does,
  title={Does multimodality improve recommender systems as expected? A critical analysis and future directions},
  author={Zhou, Hongyu and Zhang, Yinan and Sun, Aixin and Shen, Zhiqi},
  journal={arXiv preprint arXiv:2508.05377},
  year={2025}
}

@article{banham1996spatially,
  title={Spatially adaptive wavelet-based multiscale image restoration},
  author={Banham, Mark R and Katsaggelos, Aggelos K},
  journal={IEEE Trans. Image Process},
  volume={5},
  number={4},
  pages={619--634},
  year={1996},
  publisher={IEEE}
}

@article{chang2000spatially,
  title={Spatially adaptive wavelet thresholding with context modeling for image denoising},
  author={Chang, S Grace and Yu, Bin and Vetterli, Martin},
  journal={IEEE Trans. Image Process},
  volume={9},
  number={9},
  pages={1522--1531},
  year={2000},
  publisher={IEEE}
}

@article{chaovalit2011discrete,
  title={Discrete wavelet transform-based time series analysis and mining},
  author={Chaovalit, Pimwadee and Gangopadhyay, Aryya and Karabatis, George and Chen, Zhiyuan},
  journal={ACM Computing Surveys (CSUR)},
  volume={43},
  number={2},
  pages={1--37},
  year={2011},
  publisher={ACM New York, NY, USA}
}

@article{daubechies1988orthonormal,
  title={Orthonormal bases of compactly supported wavelets},
  author={Daubechies, Ingrid},
  journal={Communications on pure and applied mathematics},
  volume={41},
  number={7},
  pages={909--996},
  year={1988},
  publisher={Wiley Online Library}
}

@article{cohen1992biorthogonal,
  title={Biorthogonal bases of compactly supported wavelets},
  author={Cohen, Albert and Daubechies, Ingrid and Feauveau, J-C},
  journal={Communications on pure and applied mathematics},
  volume={45},
  number={5},
  pages={485--560},
  year={1992},
  publisher={Wiley Online Library}
}

@book{meyer1992wavelets,
  title={Wavelets and operators},
  author={Meyer, Yves},
  number={37},
  year={1992},
  publisher={Cambridge university press}
}

@article{chen2025ambiguity,
  title={Ambiguity-aware point cloud segmentation by adaptive margin contrastive learning},
  author={Chen, Yang and Duan, Yueqi and Sun, Haowen and Lu, Jiwen and Tan, Yap-Peng},
  journal={IEEE Transactions on Multimedia},
  year={2025},
  publisher={IEEE}
}

@inproceedings{chen2024adaptive,
  title={Adaptive margin contrastive learning for ambiguity-aware 3d semantic segmentation},
  author={Chen, Yang and Duan, Yueqi and Zhang, Runzhong and Tan, Yap-Peng},
  booktitle={2024 IEEE International Conference on Multimedia and Expo (ICME)},
  pages={1--6},
  year={2024},
  organization={IEEE}
}

@inproceedings{ong2023quad,
  title={Quad-tier entity fusion contrastive representation learning for knowledge aware recommendation system},
  author={Ong, Rongqing Kenneth and Qiu, Wei and Khong, Andy W. H.},
  booktitle={Proceedings of the 32nd ACM International Conference on Information and Knowledge Management},
  pages={1949--1959},
  year={2023}
}

@inproceedings{ni2025content,
  title={A content-driven micro-video recommendation dataset at scale},
  author={Ni, Yongxin and Cheng, Yu and Liu, Xiangyan and Fu, Junchen and Li, Youhua and He, Xiangnan and Zhang, Yongfeng and Yuan, Fajie},
  booktitle={Proceedings of the 34th ACM International Conference on Information and Knowledge Management},
  pages={6486--6491},
  year={2025}
}

\begin{IEEEbiography}[{\includegraphics[width=1.25in,height=1in,angle=90, keepaspectratio]{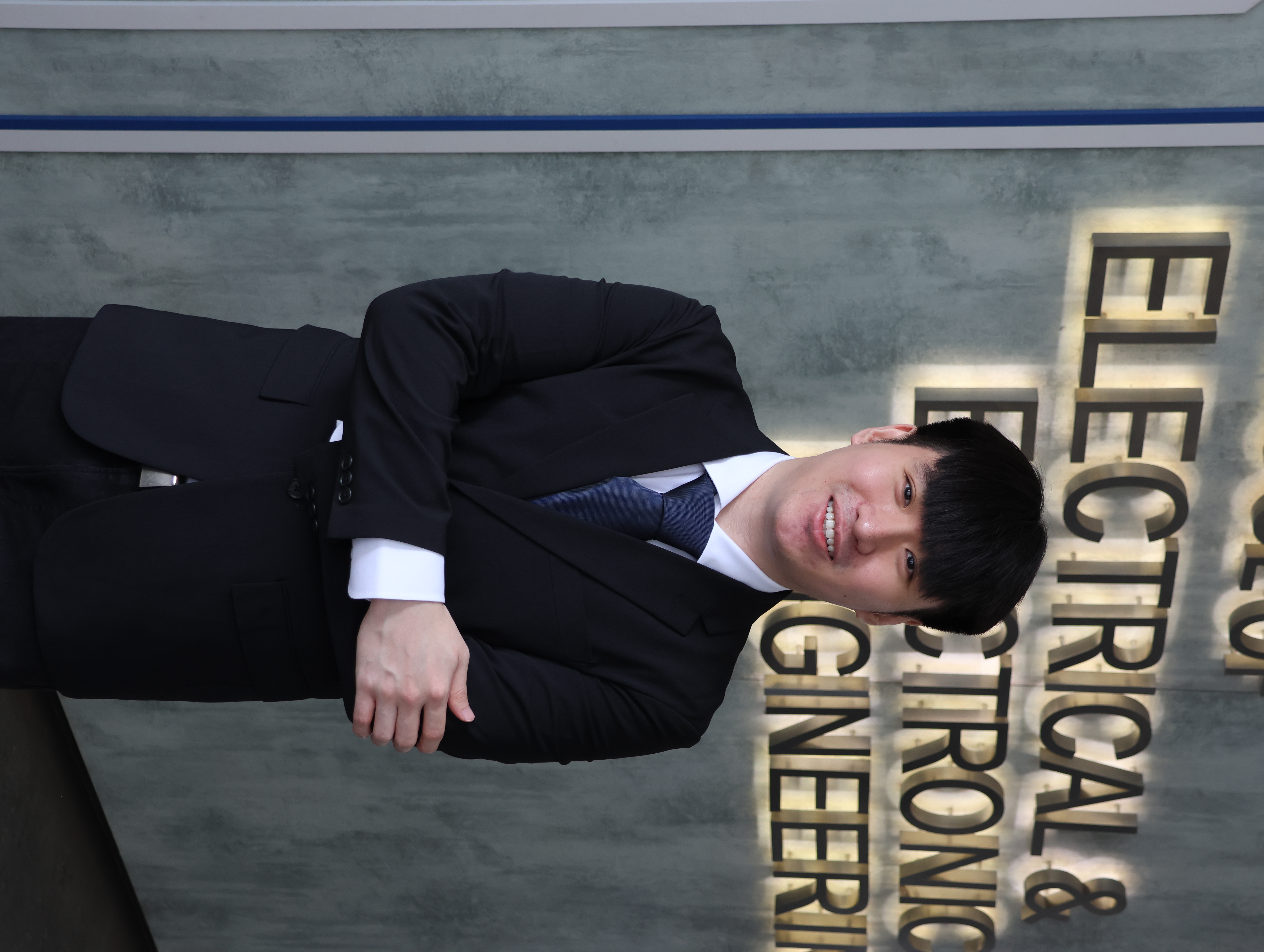}}]{Rongqing Kenneth Ong} (Member, IEEE) is currently pursuing the Doctor of Philosophy degree in the School of Electrical and Electronic Engineering at Nanyang Technological University, Singapore, supported by the Nanyang President’s Graduate Scholarship. He received the B.Eng. degree in Electrical and Electronic Engineering with Highest Distinction from NTU in 2022 under the MOE-NTU Scholarship. His research interests include information retrieval, with a focus on multi-modal and knowledge graph-enhanced recommender systems.
\end{IEEEbiography}

\begin{IEEEbiography}[{\includegraphics[width=1in,height=1.25in]{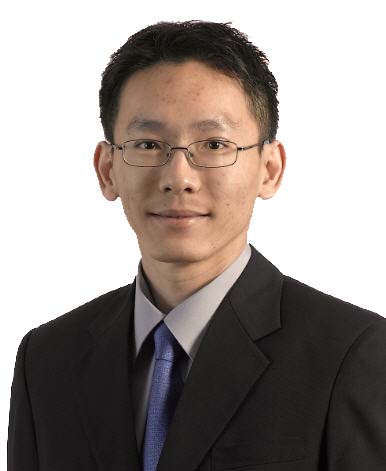}}]{Andy W. H. Khong}
(Member, IEEE) received the B.Eng. degree in electrical and electronic engineering from Nanyang Technological University, Singapore and the Ph.D. degree in signal processing from Imperial College London, London, U.K. He is currently an Associate Professor with the School of Electrical and Electronic Engineering, Nanyang Technological University, Singapore and the Lee Kong Chian School of Medicine at the same University. His research interests include human-centric machine learning techniques with applications in learning analytics, robotics, recommendation systems, and acoustic array signal processing for speech enhancement, source localization, and blind source separation. Dr. Khong had served as an Associate Editor for the IEEE/ACM TRANSACTIONS AUDIO, SPEECH, AND LANGUAGE PROCESSING and is currently an Associate Editor with the Journal of Multidimensional Systems and Signal Processing (Springer). He was a Visiting Professor with UIUC in 2012 under the Tan Chin Tuan Fellowship. Three of his papers were awarded the best student paper awards. 
\end{IEEEbiography}

\end{document}